\documentclass[12pt,preprint]{aastex}

\received{2001 Jun 7}
\accepted{2001 Aug 21}

\shorttitle{STARBURST-SEYFERT CONNECTION IN FAR-INFRARED}
\shortauthors{MOURI \& TANIGUCHI}

\begin{document}

\title{FAR-INFRARED CENSUS OF STARBURST-SEYFERT CONNECTION}

\author{HIDEAKI MOURI}
\affil{Meteorological Research Institute, Nagamine 1-1, Tsukuba 305-0052, Japan; hmouri@mri-jma.go.jp}

\and

\author{YOSHIAKI TANIGUCHI}
\affil{Astronomical Institute, Graduate School of Science, Tohoku University, Aoba, Sendai 980-8578, Japan; tani@astr.tohoku.ac.jp}

\begin{abstract}

Far-infrared flux densities are newly extracted from the {\it IRAS} database for the RSA and CfA complete samples of Seyfert galaxies. These data are used to classify the Seyfert galaxies into those where the far-infrared continuum emission is dominated by the active galactic nucleus (AGN), circumnuclear starburst, or host galaxy. 

While AGN-dominant objects consist of comparable numbers of Seyfert 1 and 2 galaxies, starburst- and host-dominant objects consist preferentially of Seyfert 2 galaxies. Thus, in addition to the dusty torus, the circumnuclear starburst region and host galaxy are important in hiding the broad-line region. 

Morphologically, starburst-dominant Seyfert galaxies are of later types and more strongly interacting than AGN-dominant Seyfert galaxies. In a later-type galaxy, the AGN central engine has a lower Eddington luminosity, and the gaseous content is higher. The gas is efficiently supplied to the starburst via a galaxy-galaxy interaction. Morphologies of host-dominant Seyfert galaxies are of various types.

Since starbursts in Seyfert galaxies are older than those in classical starburst galaxies, we propose an evolution from starburst to starburst-dominant Seyfert to host-dominant Seyfert for a late-type galaxy. An evolution from AGN-dominant Seyfert to host-dominant Seyfert is proposed for an early-type galaxy. These sequences have durations of a few $\times$ 10$^8$ yr and occur repeatedly within a galaxy during its evolution from a late type to an early type.
\end{abstract}


\section{INTRODUCTION}

The traditional classification of Seyfert galaxies is Seyfert 1 versus Seyfert 2. This is a spectroscopic discrimination. Optical broad recombination lines are seen in Seyfert 1 galaxies, while they are not seen in Seyfert 2 galaxies. The so-called torus model claims that the two types are identical but are observed in different orientations. Around the broad-line region (BLR) and the central engine, there is a parsec-scale axisymmetric structure, i.e., a dusty torus. It is claimed that the torus hides the BLR along our line of sight to a Seyfert 2 galaxy (e.g., Antonucci 1993).

However, the torus model does not explain some of differences observed between Seyfert 1 and 2 galaxies. In particular, a circumnuclear starburst is found more frequently among Seyfert 2 galaxies. The starburst region and associated gas clouds might hide the BLR in some of Seyfert 2 galaxies (e.g., Heckman et al. 1989; Osterbrock 1993; Moorwood 1996). They actually tend to have dust lanes being irregular or running across the nucleus (Malkan, Gorjian, \& Tam 1998).

The morphological type and interaction property are also different. Seyfert 2 galaxies tend to be objects of later types and more strongly interacting (e.g., Osterbrock 1993; De Robertis, Yee, \& Hayhoe 1998; Dultzin-Hacyan et al. 1999). Such objects favor starbursts as explained later. The observed differences might be due to Seyfert 2 galaxies where the BLR is hidden by a starburst region.

Therefore, circumnuclear star formation is crucial to understanding Seyfert galaxies. It is of interest to classify a sample of Seyfert galaxies into those with and without starbursts, and compare their demographics, e.g., the number ratio of Seyfert 2 to Seyfert 1 galaxies, the morphological distribution, and so on. To the extent of our knowledge, an analysis of this kind has not been carried out.\footnote{
Just before the submission of our manuscript, we realized that a similar work had been done by Storchi-Bergmann et al. (2001; see also Cid Fernandes et al. 2001). With ultraviolet to optical spectra of nuclear regions, they had classified a sample of luminous Seyfert 2 galaxies, compared the morphological type and interaction property, and obtained results that are consistent with ours.}

Since the spectral energy distribution is different between Seyfert galaxies with and without starbursts, their relative numbers in a certain sample are affected by its selection criterion. This selection bias is serious if Seyfert 1 galaxies are compared directly with Seyfert 2 galaxies, but it is not serious in the above approach.

The useful signature of a starburst is an enhancement of far-infrared (FIR) continuum emission, which is due to dust reradiation of ultraviolet photons from OB stars (e.g., Rodr\'iguez Espinosa, Rudy, \& Jones 1987; Wilson 1988; Mouri \& Taniguchi 1992). When dominated by a starburst, the FIR spectrum provides constraints on the stellar population. For a large number of Seyfert galaxies, FIR flux densities at 60 and 100 \micron\ were measured by the {\it Infrared Astronomical Satellite (IRAS)}.

We accordingly study circumnuclear star formation in Seyfert galaxies. From the raw {\it IRAS} data, FIR flux densities are newly extracted for well-defined samples of Seyfert galaxies (\S2). We also obtain FIR data of nuclear-starburst galaxies and construct a scheme to select starburst-dominant Seyfert galaxies (\S3). Using this scheme, we classify the Seyfert galaxies, compare their observational properties, and discuss mechanisms to hide a BLR as well as a connection between starburst and Seyfert activities (\S4).

Throughout this paper, we focus on Seyfert galaxies, i.e., galaxies with moderate-luminosity active galactic nuclei (AGNs). High- and low-luminosity AGNs, e.g., QSOs and LINERs, are not of our interest. They could have different connections with starburst activities. Seyfert galaxies have several subtypes. We put the subtypes 1.0 and 1.5 into Seyfert 1, and the subtypes 1.8, 1.9, and 2.0 into Seyfert 2. Seyfert $1.8+1.9$ galaxies are defined to exhibit faint broad lines. These intermediate types, which have an interesting character, are shown separately in the following diagrams.

\section{SAMPLES AND DATA}

Our Seyfert galaxies are those from optical spectroscopic surveys of two complete magnitude-limited samples of galaxies: the Palomar survey for the Revised Shapley-Ames (RSA) Catalog of Bright Galaxies and the Center for Astrophysics (CfA) redshift survey. Spectroscopic classifications were made with homogeneous data of the surveys. For reference, we also study Markarian (Mrk) starburst galaxies. These three samples of galaxies are listed along with observational data in Appendix A.

RSA galaxies serve as a fair representation of the local galaxy population. They lie in the northern sky and are equal to or brighter than 12.5 mag in the total $B$-band. Ho, Filippenko, \& Sargent (1997a) conducted an extensive survey of the nuclei and found 52 Seyfert galaxies.

CfA galaxies are more distant and luminous than RSA galaxies. The selection criteria are $\delta \ge 0\arcdeg$ and $b \ge 40\arcdeg$ or $\delta \ge -2\fdg5$ and $b \le -30\arcdeg$, and $m_B \le 14.5$. Here $m_B$ is the apparent blue magnitude on the Zwicky-$B(0)$ system for the entire galaxy. Huchra \& Burg (1992) and Osterbrock \& Martel (1993) defined a sample of 48 Seyfert galaxies.

\placetable{1}

Table 1 shows that CfA Seyfert 1 galaxies lie over larger distances than CfA Seyfert 2 galaxies. Seyfert 1 galaxies harbor bright nuclei, which contribute significantly to the total blue magnitudes (Osterbrock \& Shaw 1988). To correct for this bias, our CfA Seyfert 1 sample is restricted to galaxies with $m_B \le 14.0$. Their median distance is similar to that of CfA Seyfert 2 galaxies. We have accordingly excluded extreme objects such as the ultraluminous infrared galaxy Mrk 231. It should be noted that observational data of CfA Seyfert 1 galaxies with $m_B > 14.0$ are nevertheless given in Appendix A.

The above bias is negligible exceptionally in the RSA Seyfert sample, where the nucleus is not luminous and overwhelmed by the host galaxy in the $B$-band (Maiolino \& Rieke 1995). In addition, while the CfA sample misses galaxies with low-luminosity Seyfert nuclei (Huchra \& Burg 1992), the RSA sample is complete to such objects (Maiolino \& Rieke 1995; Ho et al. 1997a). Thus the RSA sample is more reliable than the CfA sample. 

Nevertheless, the CfA sample is complete to galaxies with luminous Seyfert nuclei, which are relatively rare in the RSA sample. It is interesting to compare these two samples. The CfA sample also consists of well-studied objects. Only with this sample, statistical analyses of some observational quantities are possible.

Mrk galaxies are those exhibiting strong ultraviolet continua. Balzano (1983) selected 102 starburst galaxies. They do not form a complete sample but have been regarded as classical examples of galaxies with ongoing nuclear starbursts. We exclude faint objects and use 57 objects in the same magnitude range as CfA Seyfert galaxies, $m_B \le 14.5$.

For the sample galaxies, FIR flux densities at 60 and 100 \micron\ are extracted from the {\it IRAS} database using the processor SCANPI, which estimates total flux densities of an extended source (Appendix A). Since galaxies are often resolved by the {\it IRAS} detectors, point-source flux densities given in the current {\it IRAS} catalogs tend to be underestimates of the true FIR emission. As a measure of FIR properties of those galaxies, the total flux densities obtained here are preferable. Even for point sources, SCANPI provides improved sensitivity and accuracy. If we exclude galaxies where the {\it IRAS} data are not available, our final samples consist of 50 RSA Seyfert, 37 CfA Seyfert, and 56 Mrk starburst galaxies.

The observational data such as the Zwicky-$B(0)$ magnitude $m_B$ and the Hubble type index $T$ are compiled from the literature (Appendix A). Although the $m_B$ values are mostly photographic, they are sufficiently accurate for our purpose (${\rm error} \simeq \pm 0.3$ mag; Huchra \& Burg 1992) and have been measured for almost all the galaxies. The other data are also accurate. This is because we have excluded faint, distant, or extreme objects from the samples.

\section{CLASSIFICATION SCHEME}

Using Mrk starburst galaxies, we construct a scheme to classify Seyfert galaxies into those where the FIR emission is dominated by the AGN, starburst, or host galaxy. The last class is not relevant directly to the starburst-Seyfert connection but exists as a definite group. If host-dominant Seyfert galaxies were not separated from AGN- and starburst-dominant Seyfert galaxies, distinction between the latter two would be unclear. Since RSA, CfA, and Mrk galaxies have different distances (Table 1), we base our classification scheme on distance-independent quantities.

Starburst-dominant Seyfert galaxies are defined as those with $F_{\rm FIR}/F_B > 1$. Here $F_{\rm FIR}$ is the FIR flux between 40 and 120 \micron\ (Fullmer \& Lonsdale 1989), and $F_B$ is the monochromatic flux at 0.43 \micron\ (Lang 1980, p. 560). Their definitions are
\begin{equation}
\label{equ1}
F_{\rm FIR} = 1.26 \times 10^{-11} \left( 2.58 S_{60} + S_{100} \right) \;
{\rm erg} \: {\rm s}^{-1} \: {\rm cm}^{-2},
\end{equation}
and
\begin{equation}
\label{equ2}
F_B = 2.16 \times 10^{-5} \: {\rm dex} \left( -0.4 m_B \right) \;
{\rm erg} \: {\rm s}^{-1} \: {\rm cm}^{-2},
\end{equation}
where $S_{\lambda}$ is the flux density at $\lambda$ \micron\ in units of Janskys. Thus $F_{\rm FIR}/F_B$ is equivalent to the FIR luminosity normalized by the total blue luminosity. Within and around a starburst region, there is a large amount of dust grains, which absorb ultraviolet photons from OB stars and reradiate them at FIR wavelengths. Mrk starburst galaxies actually exhibit $F_{\rm FIR}/F_B \gtrsim 1$ as shown in Figure 1$a$. For a sample of more than 1000 inactive galaxies, Devereux \& Hameed (1997) found $F_{\rm FIR}/F_B \lesssim 1$.

\placefigure{Fig1}

The observed $F_B$ value is dominated by stellar light of the host galaxy. Although star formation in the host galaxy depends on its morphological type (e.g., Kennicutt 1998), this dependence is not serious to the present criterion for starburst-dominant Seyfert galaxies. Our result for Mrk starburst galaxies and that of Devereux \& Hameed (1997) for inactive galaxies are independent of the morphological type. 

Host-dominant Seyfert galaxies are defined as those with $S_{100}/S_{60} > 3$. This is a constraint on the dust temperature. Outside star forming regions, the interstellar radiation is weak, the dust temperature is low, and the local value of $S_{100}/S_{60}$ is large. For example, diffuse gas in our Galaxy, the cirrus, exhibits $S_{100}/S_{60} = 4.8$ (Boulanger \& P\'erault 1988), which corresponds to the dust temperature of 27 K if dust grains radiate as a graybody with an emissivity proportional to $\lambda^{-1}$. On the other hand, Mrk starburst galaxies exhibit $S_{100}/S_{60} \lesssim 3$ as shown in Figure 1$b$.

The criterion $F_{\rm FIR}/F_B > 1$ for starburst-dominant Seyfert galaxies is consistent with the criterion $S_{100}/S_{60} > 3$ for host-dominant Seyfert galaxies. No galaxy in our samples simultaneously exhibits $F_{\rm FIR}/F_B > 1$ and $S_{100}/S_{60} > 3$. This is because, in a starburst region, the dust temperature is high and the local value of $S_{100}/S_{60}$ is small.

Finally, AGN-dominant Seyfert galaxies are defined as those with $F_{\rm FIR}/F_B \le 1$ and $S_{100}/S_{60} \le 3$. The optical to infrared continuum of an AGN is a superposition of emissions from the central engine and from gas and dust at various distances from the central engine. Empirically, the spectral energy distribution is approximated as $S_{\lambda} \propto \lambda ^{\alpha}$ with $\alpha \lesssim 1$ (e.g., Ward et al. 1987; Elvis et al. 1994), which yields $F_{\rm FIR}/F_B \lesssim 1$ and $S_{100}/S_{60} \lesssim 2$. Since the spectrum has a turnover at a FIR wavelength and the host galaxy dominates the observed $F_B$ value, we safely expect $F_{\rm FIR}/F_B \le 1$ and $S_{100}/S_{60} \le 3$ for an AGN-dominant Seyfert galaxy.

The flux densities at 25 \micron\ and 21 cm are enhanced over that at 60 \micron\ in AGNs. These enhancements have been used to select AGN-dominant Seyfert galaxies. However, there are confusing cases such that the FIR emission at 60 \micron\ is dominated by a starburst while the mid-infrared emission at 25 \micron\ and the radio emission at 21 cm are dominated by an AGN (e.g., NGC 1068; Wilson 1988). Our selection criterion is thus preferable.

\section{RESULTS AND DISCUSSION}

RSA and CfA Seyfert galaxies are classified into those dominated by the AGN, starburst, or host galaxy. Their statistics are summarized in Table 2. Their object numbers are shown as a function of morphological type in Figures 2 and 3. For these and the other histograms, open, hatched, and filled areas are used to denote Seyfert $1.0 + 1.5$, $1.8 +1.9$, and 2.0 galaxies, respectively. The classification for each of the objects is given in Appendix A.

\placetable{2}
\placefigure{Fig2}
\placefigure{Fig3}

We believe that our classification is reliable. Table 2 lists well-known Seyfert galaxies in the RSA and CfA samples. They fall into appropriate classes. For the other galaxies also, our classification is consistent with those in the literature, e.g., a catalog of Seyfert galaxies with circumnuclear starbursts (Gu, Dultzin-Hacyan, \& de Diego 2001), if we admit that the relative importance among the AGN, starburst, and host galaxy could vary for the wavelength. In addition, properties studied with our histograms are systematically different among the three classes. For essential differences, the confidence level estimated from the Kolmogorov-Smirnov test is given in the figure caption.

Figures 2 and 3 highlight our most important results. While AGN-dominant objects consist of comparable numbers of Seyfert 1 and 2 galaxies, starburst- and host-dominant objects consist preferentially of Seyfert 2 galaxies. Morphological types of AGN-dominant Seyfert galaxies are earlier than those of starburst-dominant Seyfert galaxies. Host-dominant Seyfert galaxies are of various morphologies. They are numerous in the RSA sample but not so in the CfA sample. These characteristics form the basis of the following discussion.

\subsection{Obscuration of Broad-Line Region}

Here we discuss mechanisms to hide a BLR. To estimate interstellar extinction outside the AGN, we use the narrow-line Balmer decrement $F_{{\rm H}\alpha}/F_{{\rm H}\beta}$.\footnote{
Since our Seyfert galaxies were defined on flux ratios of optical emission lines (\S2), the AGN dominates the observed H$\alpha$ and H$\beta$ fluxes even if there exists a circumnuclear starburst. Since the narrow-line region is larger than the BLR, the $F_{{\rm H}\alpha}/F_{{\rm H}\beta}$ values do not necessarily represent extinction toward the BLRs of the individual Seyfert galaxies. They are nevertheless useful in a statistical analysis of the BLR extinction.} 
Its intrinsic value is about 3. The Balmer decrement was measured for RSA Seyfert galaxies by Ho et al. (1997a). Their data are shown in Figure 4. Seyfert 1 galaxies exhibit $F_{{\rm H}\alpha}/F_{{\rm H}\beta} \le 5$, except for an edge-on galaxy NGC 4235. Only when the extinction is small, the BLR is observable.

\placefigure{Fig4}

\subsubsection{AGN-Dominant Seyfert Galaxies}

The BLR of an AGN-dominant Seyfert 2 galaxy is hidden by the dusty torus. As in the case of Seyfert 1 galaxies, an AGN-dominant Seyfert 2 galaxy always exhibits a small value of the Balmer decrement ($F_{{\rm H}\alpha}/F_{{\rm H}\beta} \le 5$; Fig. 4). Outside the AGN, there is not sufficient material to hide the BLR.

Since the dusty tori are in random orientations, their typical opening angle $\theta_{\rm torus}$ is obtained from the number fraction $f_{\rm S1}$ of Seyfert 1 galaxies as $\cos (\theta_{\rm torus}/2) = 1 - f_{\rm S1}$ (Osterbrock \& Shaw 1988; Osterbrock \& Martel 1993; see also Lawrence 1991). In our RSA and CfA samples, respectively, AGN-dominant Seyfert galaxies have $f_{\rm S1} = 0.40$ and 0.44 (Table 2), which yield $\theta_{\rm torus} \simeq 110$\arcdeg. The opening angle is unlikely to depend on the morphological type of the galaxy. This is because the morphological distribution is not different between AGN-dominant Seyfert 1 and 2 galaxies (Figs. 2 and 3).

The $\theta_{\rm torus}$ values obtained previously are underestimates. The previous studies did not exclude starburst- and host-dominant objects. They are mostly Seyfert 2 galaxies where the BLR is hidden by the circumnuclear starburst region or host galaxy (see below). Our present estimate based on AGN-dominant objects alone is more reliable.

The narrow-line region of a Seyfert galaxy is often observed to be conical or biconical with the apex at the nucleus. This ionization cone is due to collimation of the ionizing radiation, which is in turn due to shadowing by the dusty torus. However, the opening angle of the ionization cone $\theta_{\rm cone}$ is less than that of the torus $\theta_{\rm torus}$ ($\theta_{\rm cone} \simeq 40$--100\arcdeg; Wilson \& Tsvetanov 1994). The effective opening angle for the ionizing radiation is relatively small, if the torus does not have a well-defined edge. Some additional processes might be at work in shaping the ionization cones (Wilson \& Tsvetanov 1994 and references therein).

\subsubsection{Starburst-Dominant Seyfert Galaxies}

Starburst-dominant objects consist preferentially of Seyfert 2 galaxies (Figs. 2 and 3; see also Table 2). The BLR is hidden efficiently by the circumnuclear starburst region.\footnote{
The term ``circumnuclear starburst region'' in our discussion on obscuration of a BLR is used to indicate both the starburst region itself and the associated gas clouds. Although a large amount of circumnuclear gas alone is sufficient to hide the BLR, the presence of such gas leads to the occurrence of a starburst. Conversely, the occurrence of a starburst requires the presence of a large amount of gas.} 
In fact, starburst-dominant Seyfert 2 galaxies exhibit large values of the Balmer decrement ($F_{{\rm H}\alpha}/F_{{\rm H}\beta} > 5$; Fig. 4), contrasting to the case of AGN-dominant Seyfert 2 galaxies.

Within the starburst-dominant class, the orientation determines whether the object is observed as a Seyfert 1 or 2 galaxy. If our line of sight is blocked by the dusty torus or circumnuclear starburst region, the BLR is not observable. The torus is responsible for hiding the BLR in Seyfert 2 galaxies with exceptionally small values of the Balmer decrement.

The circumnuclear starburst region has a covering factor of 1/2--2/3. This is because the number fraction of Seyfert 1 galaxies among starburst-dominant objects is a factor of 2--3 lower than that among AGN-dominant objects, and also because 2/3 of starburst-dominant Seyfert galaxies exhibit $F_{{\rm H}\alpha}/F_{{\rm H}\beta} > 5$ (Fig. 4). 

Between Seyfert 1 and 2 galaxies, a difference in the reddening of the narrow-line region has long been known (e.g., Dahari \& De Robertis 1988). This difference is inconsistent with the torus model and is attributable to the presence of starburst-dominant Seyfert 2 galaxies. The extinction within the narrow-line region is not different between Seyfert 1 and 2 galaxies, judging from our result for AGN-dominant objects (Fig. 4).

\subsubsection{Host-Dominant Seyfert Galaxies}

Host-dominant objects consist preferentially of Seyfert 2 galaxies (Figs. 2 and 3; see also Table 2). In addition to obscuration by the dusty torus or circumnuclear starburst region, another mechanism hampers the detection of the BLR. Since the AGNs are weak, nuclear optical spectra of these objects are dominated by bulge stars. Even if there exist the broad lines, they are severely diluted by stellar absorption features. This mechanism is important when extinction outside the AGN is small. For $F_{{\rm H}\alpha}/F_{{\rm H}\beta} \le 5$, actually, the number fraction of Seyfert 2 galaxies among host-dominant objects is higher than those among AGN- and starburst-dominant objects (Fig. 4). In host-dominant Seyfert 2 galaxies with $F_{{\rm H}\alpha}/F_{{\rm H}\beta} > 5$, a circumnuclear starburst region might exist and hide the BLR.

Since we are among the first to define host-dominant Seyfert galaxies explicitly (see also Osterbrock 1993), we give examples. NGC 3031 is a nearby bright Seyfert 1 galaxy. Its BLR is not discernible without careful removal of the underlying starlight (e.g., Filippenko \& Sargent 1988). If distant and faint, this object would be classified as a Seyfert 2 galaxy. The other example is NGC 4395, which harbors the least luminous known Seyfert nucleus (Filippenko \& Sargent 1989).

Nicastro (2000) predicted theoretically that a BLR is intrinsically absent in some of weak AGNs. This prediction is not inconsistent with our result, although we have adopted a conservative assumption that a BLR exists in every AGN.

Host-dominant Seyfert galaxies are much more numerous in our RSA sample than they are in our CfA sample. RSA galaxies harbor less luminous AGNs than CfA galaxies (\S2). With decreasing the luminosity, an AGN becomes more liable to be overwhelmed by the host galaxy. We consider that this tendency extends to LINERs, the least-luminosity AGNs.

\subsubsection{Intermediate-Type Seyfert Galaxies}
 
The number ratio of Seyfert $1.8+1.9$ to Seyfert 2.0 galaxies is higher in our CfA sample than it is in our RSA sample (Figs. 2 and 3; see also Table 2). Especially among CfA starburst-dominant objects, Seyfert $1.8+1.9$ galaxies entirely account for the observed overabundance of Seyfert 2 galaxies. CfA Seyfert galaxies harbor luminous AGNs (\S2). Their broad lines tend to be too strong to be hidden completely.

Within the individual galaxy samples, the number ratio of Seyfert $1.8+1.9$ to Seyfert 2.0 galaxies is relatively high among starburst- and host-dominant objects. The circumnuclear starburst region or host galaxy hides the BLR less completely than the dusty torus. Starburst-dominant Seyfert $1.8+1.9$ galaxies could be in orientations where the starburst region is not so optically thick. Host-dominant Seyfert $1.8+1.9$ galaxies could be just the objects where the BLR is diluted by the stellar bulge. They make up an important fraction of host-dominant objects with $F_{{\rm H}\alpha}/F_{{\rm H}\beta} \le 5$ (Fig. 4). Among the other classes of objects with $F_{{\rm H}\alpha}/F_{{\rm H}\beta} \le 5$, Seyfert $1.8+1.9$ galaxies are rare, and Seyfert 1.0+1.5 galaxies are instead as ubiquitous as Seyfert 2.0 galaxies.

\subsection{Properties of Circumnuclear Starbursts}

Circumnuclear starbursts in Seyfert galaxies are relatively old. Their starburst ages are of order 10$^8$ yr, while those of classical starburst galaxies are of order 10$^7$ yr. This fact was first noticed by Glass \& Moorwood (1985) and Mouri \& Taniguchi (1992). Up to now, the observational evidence has been obtained for various quantities: stellar colors and absorption features at optical wavelengths (Schmitt, Storchi-Bergmann, \& Cid Fernandes 1999), stellar colors at near-infrared wavelengths (Dultzin-Hacyan \& Benitez 1994; Hunt et al. 1997), equivalent width of the Br$\gamma$ line (Oliva et al. 1995), and so on. Mouri \& Taniguchi (1992) pointed out that the age difference implies an evolution from a starburst galaxy to a starburst-dominant Seyfert galaxy.

Young star formation does not exist around an AGN. The circumnuclear starburst occurs at radii of 10$^2$--10$^3$ pc, while the fueling of the AGN occurs at radii of $\ll 10^0$ pc. Since it takes time to transport gas from the circumnuclear region to the nucleus, there is a delay between the onset of the starburst and that of the AGN activity (Moorwood 1996; Schmitt et al. 1999). Accordingly, a starburst-dominant Seyfert galaxy evolves from a young starburst galaxy.

The starburst age affects the flux density ratio $S_{100}/S_{60}$ (Mouri \& Taniguchi 2000). A starburst region consists of star clusters. In a young starburst, stellar light is absorbed by dust grains in the vicinity of the individual clusters. However, in an old starburst, stellar winds and supernova explosions have pushed away those grains, and stellar light is of longer wavelengths and travel distances before absorbed by grains. The FIR emission originates in grains that lie far from the star clusters. Since the radiation field there is weak, the grain temperature is low and the local value of $S_{100}/S_{60}$ is large.

\placefigure{Fig5}

Figure 5 shows that starburst-dominant Seyfert galaxies have larger values of $S_{100}/S_{60}$ than Mrk starburst galaxies. Although some of the Seyfert galaxies appear to suffer from the AGN component with low $S_{100}/S_{60}$, an important fraction of them has $S_{100}/S_{60} > 2$. The median value $S_{100}/S_{60} = 2.29$ was obtained for a sample of old starburst galaxies by Mouri \& Taniguchi (2000; Appendix B). These galaxies have starburst ages of order 10$^8$ yr. Thus our result is consistent with the previous estimates for starburst ages of Seyfert galaxies.

\subsection{Morphological Type and Interaction Property}

Figures 2 and 3 show that AGN-, starburst-, and host-dominant Seyfert galaxies have different distributions of the morphological type (see also Table 2). While AGN-dominant Seyfert galaxies are around the types S0/a--Sa ($T = 0$--1), starburst-dominant Seyfert galaxies are around the types Sb--Sbc ($T = 3$--4). The same result was obtained by Storchi-Bergmann et al. (2000) from an analysis of near-ultraviolet spectra of Seyfert 2 galaxies. Morphological types of host-dominant Seyfert galaxies cover those of AGN- and starburst-dominant Seyfert galaxies.

The morphological difference observed between AGN- and starburst-dominant Seyfert galaxies implies that, in a later-type galaxy, the AGN is more probable to be weak and overwhelmed by a circumnuclear starburst. The mass of the AGN central engine, a black hole, scales with that of the galaxy bulge (Ferrarese \& Merritt 2000; Gebhardt et al. 2000). Thus, in a later-type galaxy, the central engine is less massive. The Eddington luminosity is accordingly lower. In addition, a later-type galaxy has a higher content of gaseous material and is more favorable for a starburst (e.g., Roberts \& Haynes 1994; Kennicutt 1998).

\placefigure{Fig6}

The upper panel of Figure 6 shows the interaction class {\it IAC} determined by Dahari \& De Robertis (1988) for CfA Seyfert galaxies. The {\it IAC} is a one-dimensional integer scale and describes the presence of a companion and distortion in the galaxy, i.e., signatures of a galaxy-galaxy interaction: $IAC = 1$ for an isolated and symmetric galaxy, and $IAC = 6$ for a strongly interacting and distorted one. While AGN-dominant Seyfert galaxies include a larger percentage of non-interacting systems, starburst-dominant Seyfert galaxies include a larger percentage of interacting systems.

The lower panel of Figure 6 shows the amplitude $\xi_0$ of the two-point correlation function $\xi(r)$ determined by De Robertis et al. (1998) for CfA Seyfert galaxies. The function $\xi (r) = (r/r_0)^{-1.77} = \xi_0 r^{-1.77}$ gives the excess probability for finding a companion at distance $r$ from the galaxy. The search radius was 250 kpc. While AGN-dominant Seyfert galaxies tend to reside in somewhat sparse environments ($\xi_0 \le 0$), starburst-dominant Seyfert galaxies tend to reside in dense environments ($\xi_0 > 0$). This is not an artifact of the well-known morphological segregation, i.e., earlier types in denser environments (e.g., Roberts \& Haynes 1994). Starburst-dominant Seyfert galaxies are of later types than AGN-dominant Seyfert galaxies. The tendency is opposite. 

We do not have enough data for host-dominant Seyfert galaxies. Their interaction properties are expected to cover those of AGN- and starburst-dominant Seyfert galaxies as in the case of morphological types, if we rely on the following discussion (see below and \S4.4).

Since an accretion rate as low as 10$^{-2}$ $M_{\sun}$ yr$^{-1}$ is sufficient to sustain the typical luminosity 10$^{10}$ $L_{\sun}$ of a Seyfert nucleus, mechanisms necessary for transporting the gas to the nucleus could be as modest as internal processes such as bars due to gravitational instabilities of the galaxy disk (Combes 2001). This is the situation in AGN-dominant Seyfert galaxies. However, the gas consumption rate in a starburst is higher, of order 10$^0$ $M_{\sun}$ yr$^{-1}$ for 10$^{10}$ $L_{\sun}$ (e.g., Moorwood 1996; Kennicutt 1998). Hence a starburst-dominant Seyfert galaxy prefers an external process, i.e., a tidal interaction with a nearby galaxy, which is more efficient to transport a large amount of gas to the circumnuclear and nuclear regions.

Overall, whether a Seyfert galaxy is dominated by the AGN or circumnuclear starburst depends on the morphological type and interaction property. AGN- and starburst-dominant Seyfert galaxies are separated more clearly in the morphological type (Figs. 2, 3, and 6). Thus, although the relatively unclear separation in the interaction property could be partially due to a delay in the emergence of the activity after the onset of interaction, the morphological type appears to be more important. For example, if the galaxy is of a late type and thus contains a large amount of gas, even a modest mechanism supplies sufficient fuel to the starburst. The morphological type also affects consequences of an interaction. Especially in a late-type galaxy, the absence of a massive bulge enhances the gas inflow and hence the starburst (Mihos \& Hernquist 1994a).\footnote{
The exception is a major merger (Mihos \& Hernquist 1994b). Even in an early-type galaxy, merging with a galaxy of comparable mass eventually overwhelms the effect of the bulge. We do not consider this violent event, which tends to result in an extreme object such as an ultraluminous infrared galaxy.}

\subsection{Possible Evolutionary Scenario}

The evolution of Seyfert galaxies was discussed by, among others, Heckman et al. (1989), Osterbrock (1993), and Moorwood (1996). Their scenarios are summarized as, in terms of our classification scheme, starburst $\rightarrow$ starburst-dominant Seyfert $\rightarrow$ AGN-dominant Seyfert $\rightarrow$ host-dominant Seyfert $\rightarrow$ LINER. Here we propose a somewhat different scenario.

Observed ages of circumnuclear starbursts represent the typical lifetime of a starburst-dominant Seyfert galaxy, of order 10$^8$ yr (\S4.2). The lifetimes of AGN- and host-dominant Seyfert galaxies are roughly of the same order. This is because, in our RSA sample (Fig. 2 and Table 2), the number of starburst-dominant Seyfert galaxies is comparable to those of AGN- and host-dominant Seyfert galaxies.\footnote{
The numbers of AGN- and starburst-dominant Seyfert galaxies are comparable also in our CfA sample (Fig. 3 and Table 2). However, this sample misses host-dominant Seyfert galaxies and is thus unsuited to estimating their typical lifetime (\S2 and \S4.1.3).} 

The lifetimes estimated here are consistent with characteristic time scales of Seyfert and starburst activities. The Salpeter time scale is the time over which the mass of a black hole increases by a factor $e$ (e.g., Osterbrock 1993). It is of order 10$^8$ yr,  if the efficiency in converting the rest mass energy of infalling gas to the radiation is 10\% and the bolometric luminosity is 10\% of the Eddington luminosity. The gas consumption time scale is the time over which a starburst consumes the available gas at the observed star formation rate. It is of order 10$^8$ yr for classical starburst galaxies (e.g., Moorwood 1996; Kennicutt 1998; but see also Appendix B).

Our estimates imply that AGN- and starburst-dominant Seyfert galaxies do not have a direct evolutionary connection. There are systematical differences in the morphological type and interaction property (\S4.3), which do not vary markedly in $10^8$ yr. The dynamical time scale of a galaxy is of order 10$^8$ yr. It takes much longer time for a galaxy to change from a late type to an early type. The time scale of an interaction is also $\gg 10^8$ yr. In the case of an unbound interaction, for example, the companion with a typical relative velocity of 300 km s$^{-1}$ does not move a significant distance away from the Seyfert galaxy in 10$^8$ yr (De Robertis et al. 1998). Many interacting systems actually exhibit no signs of a Seyfert activity (Dultzin-Hacyan et al. 1999; see also De Robertis et al. 1998).

The other implication is that a Seyfert activity is repetitive. Since 10\% of the RSA galaxies are Seyfert galaxies (Ho et al. 1997a), every galaxy spends 10\% of the total lifetime as a Seyfert galaxy. The present-day age of a galaxy is of order $10^{10}$ yr. Hence a Seyfert activity with a duration of order $10^8$ yr has occurred several times within the same galaxy. There could even exist a possibility that an interacting galaxy becomes repeatedly a Seyfert galaxy (see also Appendix B).

Starburst-dominant Seyfert galaxies evolve from young starburst galaxies (\S4.2). These are late-type galaxies. Since starbursts occur also in early-type galaxies (Fig. 1; see also Kennicutt 1998), some of AGN-dominant Seyfert galaxies could evolve from young starburst galaxies. Their vigorous AGNs could readily overwhelm the starbursts. If a supermassive black hole does not exist at the center, a starburst galaxy evolves to an old starburst galaxy such as those studied by Mouri \& Taniguchi (2000; Appendix B). The origin of a supermassive black hole is attributable to a past starburst episode. It was recently discovered that the starburst galaxy M 82 harbors a black hole with a mass being intermediate between those of stellar-mass and supermassive black holes (Kaaret et al. 2001).

With a decrease of the fuel supply, AGN- and starburst-dominant Seyfert galaxies become less luminous and evolve to host-dominant Seyfert galaxies. Their number density is comparable to that of AGN- or starburst-dominant Seyfert galaxies at each of the morphological types (Fig. 2). With a further decrease of the fuel supply, host-dominant Seyfert galaxies evolve to LINERs.

The typical evolution of a Seyfert galaxy is summarized as follows. Suppose a late-type spiral galaxy where gas fueling into its nuclear region is triggered internally or externally. The resultant starburst leads to formation of a supermassive black hole. When the gas fueling occurs again, our galaxy evolves to starburst $\rightarrow$ starburst-dominant Seyfert $\rightarrow$ host-dominant Seyfert $\rightarrow$ LINER. This sequence is repeated a few times. The repetition of starburst and Seyfert events exhausts the gas and increases the mass of the black hole. The galaxy morphology changes to an early type. Consequently, our galaxy becomes to evolve to (starburst $\rightarrow$) AGN-dominant Seyfert $\rightarrow$ host-dominant Seyfert $\rightarrow$ LINER. Indeed some evolutionary sequences could start from middle stages.

Our scenario is in accordance with a likely evolution process of galaxy bulges, i.e., repetition of gas inflows from the disk triggered internally or externally (e.g., Wyse, Gilmore, \& Franx 1997). It is predicted that bulges of many early-type galaxies contain old stars formed during distinct starburst episodes in the past. Although the existing data are not conclusive, this prediction would provide an observational test for our scenario.

\section{CONCLUDING SUMMARY}

FIR flux densities at 60 and 100 \micron\ have been extracted from the {\it IRAS} database for the RSA and CfA complete samples of Seyfert galaxies. We have subsequently classified the Seyfert galaxies into those where the FIR continuum emission is dominated by the AGN, starburst, or host galaxy (see Table 2 for definition and number statistics). This classification has turned out to be more fundamental than the traditional spectroscopic classification, Seyfert 1 versus Seyfert 2.

AGN-dominant objects consist of comparable numbers of Seyfert 1 and 2 galaxies. The dusty torus with an opening angle of $\theta_{\rm torus} \simeq 110$\arcdeg\ is responsible for hiding the BLR. The morphological types are around S0/a--Sa (Figs. 2 and 3). There are few interacting systems (Fig. 6).

Starburst-dominant objects consist preferentially of Seyfert 2 galaxies. Obscuration of a BLR by a circumnuclear starburst region is important. Its covering factor is 1/2--2/3. The observed narrow-line Balmer decrements are actually high (Fig. 4). The morphological types are around Sb--Sbc (Figs. 2 and 3). There are many interacting systems (Fig. 6).

Host-dominant objects consist preferentially of Seyfert 2 galaxies. They are numerous in the less-luminous RSA sample. Even if the BLR is not obscured, its emission is diluted by stellar absorption features. The morphological types cover those of AGN- and starburst-dominant Seyfert galaxies (Fig. 2).

The torus model is thus incomplete. Seyfert 2 galaxies where the BLR is hidden by the torus are in the minority. It is estimated from our result for AGN-dominant objects that the total number of such Seyfert 2 galaxies is as small as the total number of Seyfert 1 galaxies.

The condition for predominance of a circumnuclear starburst is determined by the morphological type, which is related to the mass of the central black hole and the total amount of available gas, and the interaction property, which is related to the efficiency of gas fueling. The former appears to be more important than the latter.

Since starbursts in Seyfert galaxies are older than those in classical starburst galaxies (Fig. 5), we have proposed the evolutionary path of starburst $\rightarrow$ starburst-dominant Seyfert $\rightarrow$ host-dominant Seyfert $\rightarrow$ LINER for a late-type galaxy. The evolutionary path proposed for an early-type galaxy is (starburst $\rightarrow$) AGN-dominant Seyfert $\rightarrow$ host-dominant Seyfert $\rightarrow$ LINER. The individual stages of Seyfert activities have time scales of order 10$^8$ yr. These sequential activities occur several times within a galaxy during its evolution from a late type to an early type.

\acknowledgments
The processor SCANPI was developed and is maintained at the Infrared Processing and Analysis Center, California Institute of Technology. We are grateful to the referee, W. C. Keel, for helpful comments. We are also grateful for financial support from the Japanese Ministry of Education, Science, and Culture under grants 10044052 and 10304013.

\appendix
\section{OBSERVATIONAL DATA AND THEIR STATISTICS}

Table 3 summarizes observational data of RSA and CfA Seyfert galaxies. Table 4 summarizes observational data of Mrk starburst galaxies with $m_B \le 14.5$. Table 5 gives median values of several quantities for these galaxies. They were calculated with the Kaplan-Meier estimator (Feigelson \& Nelson 1985), which is implemented in the numerical code ASURV Rev. 1.2 (Lavalley, Isobe, \& Feigelson 1992). This estimator cannot handle upper and lower limits simultaneously. If necessary, we handled data with upper and lower limits separately, and calculated the object-number-weighted average.

\placetable{3}
\placetable{4}
\placetable{5}

The present sample of Mrk starburst galaxies does not include Mrk 732 and Mrk 897, which are in the original catalog (Balzano 1983) but have turned out to be Seyfert galaxies (e.g., Osterbrock \& Pogge 1987; Keel 1996).

For RSA and CfA Seyfert galaxies, respectively, the spectral types 1.0, 1.5, 1.8, 1.9, and 2.0 were from Ho et al. (1997a) and Osterbrock \& Martel (1993). When these two references assign different types to the same object, we preferred the type in Osterbrock \& Martel (1993).

The coordinates were mainly from the Third Reference Catalogue of Bright Galaxies (RC3) compiled by de Vaucouleurs et al. (1991; see also Corwin, Buta, \& de Vaucouleurs 1994). The coordinates of the Seyfert galaxies Mrk 205 and Mrk 841 were from Huchra \& Burg (1992). Those of the starburst galaxies Mrk 220, Mrk 409, and Mrk 489 were determined with the 21 cm image of the NRAO VLA Sky Survey (Condon et al. 1998). For these coordinates, we estimated the {\it IRAS} fluxes.

The distances were mainly from the nearby galaxy catalog of Tully (1988), which is almost complete to galaxies closer than 40 Mpc. For galaxies that are not listed in Tully (1988), the distance was estimated from the recession velocity in the cosmic microwave background frame and a Hubble constant $H_0 = 75$ km s$^{-1}$ Mpc$^{-1}$. The recession velocities were from RC3. Exceptionally, those of the Seyfert galaxies Mrk 205 and Mrk 841 were from Huchra et al. (1983).

The Hubble type indices $T$ were mainly from RC3. For the Seyfert galaxies Mrk 335, I Zw 1, NGC 1144, Mrk 461, NGC 5695, and NGC 6104 as well as the starburst galaxies Mrk 102, Mrk 190, Mrk 203, Mrk 307, Mrk 353, Mrk 691, Mrk 809, and Mrk 1002, the indices were from catalogs of the CfA survey (Huchra, Vogeley, \& Geller 1999; Huchra, Geller, \& Corwin 1995; Huchra et al. 1983).

The apparent blue magnitudes $m_B$ on the Zwicky-$B(0)$ system were from, in decreasing order of preference, Huchra et al. (1999, 1995, 1990, 1983), Zwicky et al. (1961--1968), and Markarian et al. (1989). For the Seyfert galaxies NGC 1358 and NGC 1667, we used the RC3 photographic magnitude.

The {\it IRAS} flux densities at 12, 25, 60, and 100 \micron\ were determined using the processor SCANPI, which projects the individual scan tracks crossing the target position onto a mean one-dimensional scan, subtracts baselines, coadds the scans, and provides estimates of flux densities. There are two coaddition methods. We usually adopted the noise-weighted mean. If it disagrees with the median by more than 1$\sigma$, we adopted the latter. Then we examined the scan profile and judged whether the source was unresolved or resolved. For an unresolved source, we used the best-fitting point-source template as the estimator of the flux density. If the fit was poor, we instead used the peak flux density for a bright source or the flux density integrated over a fixed range for a faint source. The integration ranges at 12, 25, 60, and 100 \micron\ were $\pm 2\arcmin$, $\pm 2\arcmin$, $\pm 2\farcm5$, and $\pm 4\arcmin$, respectively. These ranges were also used to integrate the flux density of a faint resolved source. For a bright resolved source, the integration was made between the zero crossings. The upper limit was set to be $3\sigma$. If there could exist contamination of a nearby object, we indicate so in Table 3 or 4. Flux densities were determined in the same manner by Soifer et al. (1989) and Sanders et al (1995). If available, we used their data. For galaxies with large apparent diameters, we preferentially used two-dimensionally coadded flux densities of Rice et al. (1988).

\section{COMMENTS ON NUCLEAR-STARBURST GALAXIES}

The star formation rate of a nuclear starburst varies largely in time. A plausible approximation is a series of burst events. The individual bursts have durations of order 10$^6$--10$^7$ yr and probably correspond to formation of (groups of) star clusters. The entire series could have a long duration, $\gtrsim$ 10$^8$ yr, which reflects time scales of gas fueling, gas consumption due to star formation, gas removal due to stellar winds and supernova explosions, and so on (e.g., Leitherer 2001).

Classical starburst galaxies such as Mrk objects are biased in favor of galaxies with ongoing bursts, i.e., young starburst galaxies. They are selected by optical line emission, which originates in ionized gas around O stars. However, there are numerous B stars also, which have long lifetimes and are responsible for most of the FIR emission. With an increase of the burst age above the burst duration, O stars become few, but B stars remain numerous, and the FIR emission remains strong. Mouri \& Taniguchi (2000) studied such old starburst galaxies in the RSA sample. These galaxies are at quiescent stages between burst events or at post-starburst stages (see also Kennicutt 1998).

A nuclear starburst occurs several times within the same galaxy. Mrk starburst galaxies have burst ages of order $10^7$ yr and account for 3\% of all the galaxies (Balzano 1983; Calzetti 1997). Old starburst galaxies studied by Mouri \& Taniguchi (2000) have burst ages of order 10$^8$ yr and account for 30\%--40\% of all the galaxies. These number fractions are too large to be explained by a single series of burst events in a galaxy, which has an age of order 10$^{10}$ yr. The prototypical starburst galaxy M 82 harbors a fossil starburst $6 \times 10^8$ yr old, which had a duration of a few $\times$ 10$^8$ yr, and an ongoing starburst 1--$2 \times 10^7$ yr old (de Grijs, O'Connell, \& Gallagher 2001). Since a starburst is caused by gas fueling into the nuclear region as in the case of a Seyfert activity, this fact is consistent with our scenario that the gas fueling and hence the Seyfert activity are repetitive (\S4.4).

Morphological types of old starburst galaxies tend to be later than those of young starburst galaxies and Seyfert galaxies (Mouri \& Taniguchi 2000; see also Ho, Filippenko, \& Sargent 1997b). Since a supermassive black hole prefers an early type (\S4.3), this tendency is consistent with our scenario that a young starburst galaxy without a supermassive black hole evolves to an old starburst galaxy (\S4.4).


\clearpage

\begin{deluxetable}{lcc}

\tabletypesize{\scriptsize}
\tablenum{1}
\tablecolumns{3}
\tablewidth{0pc}
\tablecaption{SAMPLE CHARACTERISTICS}

\tablehead{
\colhead{Sample} &
\colhead{Number} &
\colhead{Distance} \\
\colhead{} &
\colhead{} &
\colhead{Mpc}}

\startdata
RSA Seyfert 1                  & \phn9 & \phn20.6 \\
RSA Seyfert 2                  &    43 & \phn19.5 \\
CfA Seyfert 1 $(m_B \le 14.5)$ &    22 &    107.3 \\
CfA Seyfert 1 $(m_B \le 14.0)$ &    13 & \phn58.4 \\
CfA Seyfert 2                  &    26 & \phn64.7 \\
Mrk starburst $(m_B \le 14.5)$ &    57 & \phn42.5 
\enddata

\tablecomments{For the distance, we give the median value.}  

\end{deluxetable}


\begin{deluxetable}{llcccl}

\tabletypesize{\scriptsize}
\setlength{\tabcolsep}{0.045in}
\tablenum{2}
\tablecolumns{6}
\tablewidth{0pc}
\tablecaption{FIR CLASSIFICATION OF SEYFERT GALAXIES}

\tablehead{
\colhead{Class} &
\colhead{Definition} &
\colhead{Sample} &
\colhead{Number} &
\colhead{$T$} &
\colhead{Example} \\ }

\startdata
AGN Dominant       & $F_{\rm FIR}/F_B \le 1$ and $S_{100}/S_{60} \le 3$ & RSA & 6 : 1 :\phn8  & 0.0     & NGC 4051, NGC 4151, NGC 5548 \\
                   &                                                    & CfA & 8 : 3 :\phn7  & 1.0     & NGC 4051, NGC 4151, NGC 5548 \\
Starburst Dominant & $F_{\rm FIR}/F_B > 1$                              & RSA & 2 : 3 :\phn9  & 4.0     & NGC 1068, NGC 1275, NGC 3227 \\
                   &                                                    & CfA & 4 : 7 :\phn6  & 3.0     & NGC 1068, NGC 3227, NGC 7469 \\
Host Dominant      & $S_{100}/S_{60} > 3$                               & RSA & 1 : 5 :12     & 2.5     & NGC 3031, NGC 4395, NGC 4501 \\
                   &                                                    & CfA & 1 : 1 :\phn0  & \nodata & NGC 4395 \\
\enddata

\tablecomments{The galaxy numbers are given for the individual spectroscopic types ($1.0+1.5 : 1.8+1.9 : 2.0$). For the Hubble type index $T$, the median value is given. CfA Seyfert $1.0+1.5$ galaxies with $m_B > 14.0$ have been excluded from the statistics.}

\end{deluxetable}


\begin{deluxetable}{lcccrrrrrlrrrrc}

\tabletypesize{\scriptsize}
\setlength{\tabcolsep}{0.02in}
\tablenum{3}
\tablecolumns{15}
\tablewidth{0pc}
\tablecaption{RSA AND CFA SEYFERT GALAXIES}

\tablehead{
\colhead{Name} &
\colhead{Sample} &
\multicolumn{2}{c}{Class} &
\colhead{$\alpha$(1950)} &
\colhead{$\delta$(1950)} &
\colhead{$V$} &
\colhead{$D$} &
\colhead{$T$} &
\colhead{$m_B$} &
\colhead{$S_{12}$} &
\colhead{$S_{25}$} &
\colhead{$S_{60}$} &
\colhead{$S_{100}$} &
\colhead{Ref.} \\
\colhead{} &
\colhead{} &
\colhead{Op.} &
\colhead{FIR} &
\colhead{h m s} &
\colhead{d m s} &
\colhead{km/s} &
\colhead{Mpc} &
\colhead{} &
\colhead{mag} &
\colhead{Jy} &
\colhead{Jy} &
\colhead{Jy} &
\colhead{Jy} &
\colhead{} \\
\colhead{(1)} &
\colhead{(2)} &
\colhead{(3)} &
\colhead{(4)} &
\colhead{(5)} &
\colhead{(6)} &
\colhead{(7)} &
\colhead{(8)} &
\colhead{(9)} &
\colhead{(10)} &
\colhead{(11)} &
\colhead{(12)} &
\colhead{(13)} &
\colhead{(14)} &
\colhead{(15)} }

\startdata
Mrk 334                   & \phm{R }C & 1.8 & S       & 00 00 35.5 &    21 40 53 &  6261 & \nodata &    99.0 &     14.4  &    0.19$\pm$0.03\phm{:} &    1.05$\pm$0.04\phm{:} &    3.92$\pm$0.03\phm{:} &    4.32$\pm$0.23\phm{:} & 1\\
Mrk 335                   & \phm{R }C & 1.0 & A       & 00 03 45.1 &    19 55 27 &  7347 & \nodata &  $-$6.0 &     14.0  &    0.24$\pm$0.03\phm{:} &    0.44$\pm$0.03\phm{:} &    0.35$\pm$0.04\phm{:} & $< 0.37$:\phs\phm{0.00} & 1\\
NGC 185                   & R\phm{ C} & 2.0 & H       & 00 36 12.0 &    48 03 50 &  -502 &  0.7    &  $-$5.0 &     11.0  &    0.04\phs\phm{0.00:}  & $< 0.03$\phs\phm{0.00:} &    0.31\phs\phm{0.00:}  &    1.93\phs\phm{0.00:}  & 2\\ 
A0048$+$29                & \phm{R }C & 1.0 & S       & 00 48 52.9 &    29 07 45 & 10475 & \nodata &     3.0 &     14.5  &    0.12$\pm$0.03\phm{:} &    0.15$\pm$0.04\phm{:} &    1.07$\pm$0.03\phm{:} &    1.88$\pm$0.19\phm{:} & 1\\
I Zw 1                    & \phm{R }C & 1.0 & S       & 00 50 57.9 &    12 25 23 & 17978 & \nodata &     5.0 &     14.0  &    0.50$\pm$0.02\phm{:} &    1.17$\pm$0.07\phm{:} &    2.30$\pm$0.04\phm{:} &    2.89$\pm$0.08\phm{:} & 1\\
Mrk 993                   & \phm{R }C & 1.5 & H       & 01 22 42.7 &    31 52 35 &  4382 & \nodata &     1.0 &     14.0  &    0.12$\pm$0.03\phm{:} & $< 0.09$\phs\phm{0.00:} &    0.37$\pm$0.05\phm{:} &    1.60$\pm$0.17\phm{:} & 1\\
Mrk 573                   & \phm{R }C & 2.0 & A       & 01 41 22.7 &    02 05 54 &  4817 & \nodata &  $-$1.0 &     14.0  &    0.35$\pm$0.07\phm{:} &    0.94$\pm$0.03\phm{:} &    1.08$\pm$0.04\phm{:} &    1.41$\pm$0.14\phm{:} & 1\\
NGC 676                   & R\phm{ C} & 2.0 & A       & 01 46 20.6 &    05 39 35 &  1225 & 19.5    &     0.0 &     10.5  &    0.20$\pm$0.04:       & $< 0.19$\phs\phm{0.00:} &    0.36$\pm$0.05:       &    0.96$\pm$0.13\phm{:} & 1\\
0152$+$06                 & \phm{R }C & 1.9 & A       & 01 52 44.8 &    06 22 00 &  4887 & \nodata &     3.0 &     14.5  & $< 0.08$\phs\phm{0.00:} &    0.21$\pm$0.04\phm{:} &    0.53$\pm$0.04\phm{:} &    1.25$\pm$0.32\phm{:} & 1\\
NGC 777                   & R\phm{ C} & 2.0 & \nodata & 01 57 21.1 &    31 11 16 &  4734 & \nodata &  $-$5.0 &     13.09 & $< 0.11$\phs\phm{0.00:} & $< 0.11$\phs\phm{0.00:} & $< 0.11$\phs\phm{0.00:} & $< 0.26$\phs\phm{0.00:} & 1\\
NGC 863                   & \phm{R }C & 1.0 & A       & 02 12 00.6 & $-$00 59 57 &  7660 & \nodata &     1.3 &     14.0  &    0.22$\pm$0.04\phm{:} &    0.13$\pm$0.03\phm{:} &    0.64$\pm$0.03\phm{:} &    1.20$\pm$0.15\phm{:} & 1\\
NGC 1058                  & R\phm{ C} & 2.0 & H       & 02 40 23.2 &    37 07 48 &   318 &  9.1    &     5.0 &     11.8  &    0.28$\pm$0.01\phm{:} &    0.46$\pm$0.03\phm{:} &    3.26$\pm$0.05\phm{:} &   10.60$\pm$0.15\phm{:} & 1\\
NGC 1068                  & R C       & 2.0 & S       & 02 40 06.5 & $-$00 13 32 &   911 & 14.4    &     3.0 & \phn 9.81 &   36.10$\pm$0.06\phm{:} &   84.25$\pm$0.19\phm{:} &  181.95$\pm$0.10\phm{:} &  235.87$\pm$0.22\phm{:} & 3\\  
NGC 1144\tablenotemark{a} & \phm{R }C & 2.0 & S       & 02 52 38.5 & $-$00 23 07 &  8438 & \nodata &    10.0 &     13.6  &    0.27$\pm$0.03\phm{:} &    0.58$\pm$0.03\phm{:} &    5.06$\pm$0.04\phm{:} &   11.45$\pm$0.20\phm{:} & 3\\
NGC 1167                  & R\phm{ C} & 2.0 & H       & 02 58 35.3 &    35 00 31 &  4768 & \nodata &  $-$3.0 &     14.0  & $< 0.30$:\phs\phm{0.00} & $< 0.11$:\phs\phm{0.00} & $< 0.45$:\phs\phm{0.00} &    2.04$\pm$0.21\phm{:} & 1\\
NGC 1275                  & R\phm{ C} & 1.5 & S       & 03 16 29.9 &    41 19 55 &  5104 & \nodata &    99.0 &     12.44 &    0.99$\pm$0.02\phm{:} &    3.58$\pm$0.03\phm{:} &    7.40$\pm$0.07\phm{:} &    6.90$\pm$0.44\phm{:} & 4\\
NGC 1358                  & R\phm{ C} & 2.0 & A       & 03 31 10.8 & $-$05 15 24 &  3868 & \nodata &     0.0 &     12.76 & $< 0.08$\phs\phm{0.00:} &    0.12$\pm$0.03\phm{:} &    0.32$\pm$0.04\phm{:} & $< 0.84$\phs\phm{0.00:} & 1\\
NGC 1667                  & R\phm{ C} & 2.0 & S       & 04 46 10.4 & $-$06 24 26 &  4511 & \nodata &     5.0 &     13.0  &    0.59$\pm$0.03\phm{:} &    0.70$\pm$0.04\phm{:} &    6.24$\pm$0.04\phm{:} &   16.54$\pm$0.18\phm{:} & 3\\
NGC 2273                  & R\phm{ C} & 2.0 & S       & 06 45 37.6 &    60 54 13 &  1877 & 28.4    &     0.5 &     12.5  &    0.48$\pm$0.03\phm{:} &    1.35$\pm$0.02\phm{:} &    6.59$\pm$0.05\phm{:} &    9.89$\pm$0.13\phm{:} & 4\\
NGC 2639                  & R\phm{ C} & 1.9 & H       & 08 40 03.1 &    50 23 14 &  3434 & \nodata &     1.0 &     12.4  &    0.21$\pm$0.02\phm{:} &    0.30$\pm$0.02\phm{:} &    1.98$\pm$0.02\phm{:} &    7.10$\pm$0.11\phm{:} & 1\\
NGC 2655                  & R\phm{ C} & 2.0 & A       & 08 49 09.1 &    78 24 53 &  1427 & 24.4    &     0.0 &     10.8  &    0.33$\pm$0.03\phm{:} &    0.41$\pm$0.02\phm{:} &    1.81$\pm$0.03\phm{:} &    5.03$\pm$0.05\phm{:} & 1\\
NGC 2685                  & R\phm{ C} & 2.0 & H       & 08 51 41.3 &    58 55 30 &  1003 & 16.2    &  $-$1.0 &     12.1  &    0.12$\pm$0.03\phm{:} &    0.14$\pm$0.04\phm{:} &    0.51$\pm$0.04\phm{:} &    1.66$\pm$0.10\phm{:} & 1\\
NGC 3031                  & R\phm{ C} & 1.5 & H       & 09 51 27.7 &    69 18 13 &    48 &  1.4    &     2.0 & \phn 7.88 &    5.86\phs\phm{0.00:}  &    5.42\phs\phm{0.00:}  &   44.73\phs\phm{0.00:}  &  174.02\phs\phm{0.00:}  & 2\\
NGC 3079                  & R\phm{ C} & 2.0 & S       & 09 58 35.4 &    55 55 11 &  1285 & 20.4    &     7.0 &     11.43 &    2.62$\pm$0.04\phm{:} &    3.58$\pm$0.04\phm{:} &   50.17$\pm$0.05\phm{:} &  103.40$\pm$0.15\phm{:} & 3\\
NGC 3080                  & \phm{R }C & 1.0 & A       & 09 57 14.5 &    13 17 02 & 10951 & \nodata &     1.0 &     14.5  & $< 0.07$\phs\phm{0.00:} &    0.15$\pm$0.05\phm{:} &    0.51$\pm$0.04\phm{:} &    1.29$\pm$0.09\phm{:} & 1\\
NGC 3147                  & R\phm{ C} & 2.0 & H       & 10 12 39.4 &    73 39 02 &  2875 & 40.9    &     4.0 &     11.3  &    0.93$\pm$0.03\phm{:} &    1.08$\pm$0.02\phm{:} &    8.40$\pm$0.04\phm{:} &   29.96$\pm$0.23\phm{:} & 3\\
NGC 3185                  & R\phm{ C} & 2.0 & A       & 10 14 53.2 &    21 56 20 &  1539 & 21.3    &     1.0 &     13.23 &    0.18$\pm$0.03\phm{:} &    0.21$\pm$0.05\phm{:} &    1.56$\pm$0.03\phm{:} &    3.50$\pm$0.10\phm{:} & 1\\
NGC 3227                  & R C       & 1.5 & S       & 10 20 47.6 &    20 07 00 &  1472 & 20.6    &     1.0 &     11.75 &    0.93$\pm$0.03\phm{:} &    1.85$\pm$0.06\phm{:} &    8.32$\pm$0.03\phm{:} &   18.44$\pm$0.15\phm{:} & 3\\
NGC 3254                  & R\phm{ C} & 2.0 & H       & 10 26 31.4 &    29 44 50 &  1613 & 23.6    &     4.0 &     12.41 & $< 0.10$\phs\phm{0.00:} &    0.52$\pm$0.05\phm{:} &    0.73$\pm$0.04\phm{:} &    2.60$\pm$0.07\phm{:} & 1\\
NGC 3362\tablenotemark{b} & \phm{R }C & 2.0 & \nodata & 10 42 15.2 &    06 51 28 &  8666 & \nodata &     5.0 &     13.6  &    \nodata\phm{0...}    &    \nodata\phm{0...}    &    \nodata\phm{0...}    &    \nodata\phm{0...}    & \nodata\\
NGC 3486                  & R\phm{ C} & 2.0 & A       & 10 57 40.0 &    29 14 40 &   976 &  7.4    &     5.0 &     11.10 &    0.62$\pm$0.04\phm{:} &    0.32$\pm$0.05\phm{:} &    6.24$\pm$0.04\phm{:} &   15.87$\pm$0.09\phm{:} & 3\\
A1058$+$45                & \phm{R }C & 2.0 & A       & 10 58 42.8 &    45 55 25 &  8944 & \nodata &     1.0 &     14.1  &    0.16$\pm$0.03\phm{:} &    0.33$\pm$0.03\phm{:} &    0.65$\pm$0.03\phm{:} &    1.36$\pm$0.08\phm{:} & 1\\
NGC 3516                  & R C       & 1.5 & A       & 11 03 22.7 &    72 50 25 &  2696 & 38.9    &  $-$2.0 &     12.86 &    0.40$\pm$0.02\phm{:} &    0.96$\pm$0.02\phm{:} &    2.01$\pm$0.03\phm{:} &    2.64$\pm$0.11:       & 1\\
NGC 3735                  & R\phm{ C} & 2.0 & S       & 11 33 04.8 &    70 48 42 &  2780 & 41.0    &     5.0 &     12.60 &    0.68$\pm$0.04\phm{:} &    1.17$\pm$0.03\phm{:} &    7.37$\pm$0.03\phm{:} &   19.54$\pm$0.10\phm{:} & 3\\
NGC 3786\tablenotemark{b} & \phm{R }C & 1.8 & \nodata & 11 37 04.7 &    32 11 13 &  3006 & 41.6    &     1.0 &     13.5  &    \nodata\phm{0...}    &    \nodata\phm{0...}    &    \nodata\phm{0...}    &    \nodata\phm{0...}    & \nodata\\
NGC 3941\tablenotemark{b} & R\phm{ C} & 2.0 & \nodata & 11 50 19.4 &    37 15 55 &  1183 & 18.9    &  $-$2.0 &     11.66 &    \nodata\phm{0...}    &    \nodata\phm{0...}    &    \nodata\phm{0...}    &    \nodata\phm{0...}    & \nodata\\
NGC 3976                  & R\phm{ C} & 2.0 & H       & 11 53 23.0 &    07 01 40 &  2845 & 37.7    &     3.0 &     12.27 &    0.20$\pm$0.04\phm{:} &    0.40$\pm$0.04\phm{:} &    1.97$\pm$0.04\phm{:} &    6.79$\pm$0.10\phm{:} & 1\\
NGC 3982                  & R C       & 1.9 & S       & 11 53 52.3 &    55 24 10 &  1277 & 17.0    &     3.0 &     12.45 &    0.53$\pm$0.02\phm{:} &    0.89$\pm$0.03\phm{:} &    7.35$\pm$0.04\phm{:} &   17.36$\pm$0.13\phm{:} & 3\\
NGC 4051                  & R C       & 1.0 & A       & 12 00 36.4 &    44 48 35 &   945 & 17.0    &     4.0 &     11.23 &    1.40$\pm$0.03\phm{:} &    2.19$\pm$0.04\phm{:} &   10.38$\pm$0.05\phm{:} &   23.91$\pm$0.07\phm{:} & 3\\
NGC 4138\tablenotemark{b} & R\phm{ C} & 1.9 & \nodata & 12 06 59.3 &    43 57 57 &  1151 & 17.0    &  $-$1.0 &     12.80 &    \nodata\phm{0...}    &    \nodata\phm{0...}    &    \nodata\phm{0...}    &    \nodata\phm{0...}    & \nodata\\
NGC 4151                  & R C       & 1.5 & A       & 12 08 01.6 &    39 41 09 &  1238 & 20.3    &     2.0 &     11.48 &    2.07$\pm$0.05\phm{:} &    4.97$\pm$0.05\phm{:} &    6.28$\pm$0.07\phm{:} &    7.93$\pm$0.13\phm{:} & 1\\
NGC 4168                  & R\phm{ C} & 1.9 & \nodata & 12 09 43.5 &    13 29 05 &  2614 & 16.8    &  $-$5.0 &     12.77 & $< 0.10$:\phs\phm{0.00} & $<$0.12:\phs\phm{0.00}  & $< 0.14$:\phs\phm{0.00} & $< 0.39$:\phs\phm{0.00} & 1\\
NGC 4169\tablenotemark{a} & R\phm{ C} & 2.0 & H       & 12 09 47.1 &    29 27 30 &  4093 & \nodata &  $-$2.0 &     12.9  &    0.10$\pm$0.03\phm{:} &    0.28$\pm$0.04:       &    1.17$\pm$0.05\phm{:} &    5.45$\pm$0.11\phm{:} & 1\\
NGC 4235                  & R C       & 1.0 & A       & 12 14 35.7 &    07 28 11 &  2743 & 35.1    &     1.0 &     12.86 & $< 0.08$:\phs\phm{0.00} &    0.44$\pm$0.06\phm{:} &    0.35$\pm$0.04\phm{:} &    0.88$\pm$0.11\phm{:} & 1\\
NGC 4253                  & \phm{R }C & 1.5 & S       & 12 15 55.5 &    30 05 27 &  4101 & \nodata &     1.0 &     13.7  &    0.41$\pm$0.03\phm{:} &    1.61$\pm$0.05\phm{:} &    4.38$\pm$0.04\phm{:} &    4.18$\pm$0.10\phm{:} & 1\\
NGC 4258                  & R\phm{ C} & 1.9 & H       & 12 16 29.7 &    47 34 55 &   657 &  6.8    &     4.0 & \phn 9.19 &    2.25\phs\phm{0.00:}  &    2.81\phs\phm{0.00:}  &   21.60\phs\phm{0.00:}  &   78.39\phs\phm{0.00:}  & 2\\
Mrk 205\tablenotemark{c}  & \phm{R }C & 1.0 & \nodata & 12 19 33.5 &    75 35 15 & 21458 & \nodata & \nodata &     14.5  &    \nodata\phm{0...}    &    \nodata\phm{0...}    &    \nodata\phm{0...}    &    \nodata\phm{0...}    & \nodata\\
NGC 4378                  & R\phm{ C} & 2.0 & A       & 12 22 44.3 &    05 12 13 &  2889 & 35.1    &     1.0 &     12.34 & $< 0.14$\phs\phm{0.00:} & $< 0.28$\phs\phm{0.00:} &    0.75$\pm$0.05\phm{:} &    2.01$\pm$0.10\phm{:} & 1\\
NGC 4388                  & R C       & 2.0 & S       & 12 23 14.8 &    12 56 18 &  2843 & 16.8    &     3.0 &     12.69 &    1.06$\pm$0.03\phm{:} &    3.42$\pm$0.07\phm{:} &   10.05$\pm$0.03\phm{:} &   17.40$\pm$0.18\phm{:} & 3\\
NGC 4395                  & R C       & 1.8 & H       & 12 23 20.9 &    33 49 22 &   585 &  3.6    &     9.0 &     10.84 & $< 0.11$\phs\phm{0.00:} &    0.17\phs\phm{0.00:}  &    4.21\phs\phm{0.00:}  &   12.90\phs\phm{0.00:}  & 2\\
NGC 4472                  & R\phm{ C} & 2.0 & \nodata & 12 27 13.9 &    08 16 32 &  1246 & 16.8    &  $-$5.0 & \phn 9.84 & $< 0.13$\phs\phm{0.00:} & $< 0.21$\phs\phm{0.00:} & $< 0.19$\phs\phm{0.00:} & $< 0.48$\phs\phm{0.00:} & 2\\
NGC 4477                  & R\phm{ C} & 2.0 & A       & 12 27 30.7 &    13 54 45 &  1670 & 16.8    &  $-$2.0 &     11.62 &    0.27$\pm$0.03\phm{:} &    0.39$\pm$0.04\phm{:} &    0.59$\pm$0.05\phm{:} &    0.82$\pm$0.14\phm{:} & 1\\
NGC 4501                  & R\phm{ C} & 2.0 & H       & 12 29 28.2 &    14 41 50 &  2598 & 16.8    &     3.0 &     10.49 &    2.13$\pm$0.03\phm{:} &    2.95$\pm$0.08\phm{:} &   17.56$\pm$0.08\phm{:} &   63.65$\pm$0.17\phm{:} & 3\\
NGC 4565                  & R\phm{ C} & 1.9 & H       & 12 33 51.9 &    26 15 35 &  1513 &  9.7    &     3.0 &     10.61 &    1.53\phs\phm{0.00:}  &    1.70\phs\phm{0.00:}  &    9.83\phs\phm{0.00:}  &   47.23\phs\phm{0.00:}  & 2\\
NGC 4579                  & R\phm{ C} & 1.9 & H       & 12 35 12.6 &    12 05 40 &  1843 & 16.8    &     3.0 &     10.72 &    1.11$\pm$0.05\phm{:} &    0.76$\pm$0.06\phm{:} &    5.85$\pm$0.06\phm{:} &   20.86$\pm$0.27\phm{:} & 3\\
NGC 4639                  & R\phm{ C} & 1.0 & A       & 12 40 21.7 &    13 31 56 &  1322 & 16.8    &     4.0 &     12.50 &    0.17$\pm$0.04:       &    0.11$\pm$0.03\phm{:} &    1.77$\pm$0.04\phm{:} &    5.27$\pm$0.09\phm{:} & 1\\
NGC 4698                  & R\phm{ C} & 2.0 & H       & 12 45 51.8 &    08 45 37 &  1321 & 16.8    &     2.0 &     11.81 &    0.25$\pm$0.03\phm{:} & $< 0.19$\phs\phm{0.00:} &    0.64$\pm$0.04\phm{:} &    2.05$\pm$0.12\phm{:} & 1\\
NGC 4725                  & R\phm{ C} & 2.0 & H       & 12 47 59.9 &    25 46 20 &  1487 & 12.4    &     2.0 &     10.21 &    0.32\phs\phm{0.00:}  &    0.20\phs\phm{0.00:}  &    4.18\phs\phm{0.00:}  &   20.79\phs\phm{0.00:}  & 2\\
Mrk 231                   & \phm{R }C & 1.0 & S       & 12 54 05.0 &    57 08 37 & 12447 & \nodata &     5.0 &     14.1  &    1.93$\pm$0.02\phm{:} &    8.80$\pm$0.04\phm{:} &   35.40$\pm$0.04\phm{:} &   32.28$\pm$0.12\phm{:} & 3\\
NGC 5033                  & R C       & 1.9 & S       & 13 11 09.7 &    36 51 30 &  1107 & 18.7    &     5.0 &     10.85 &    1.38\phs\phm{0.00:}  &    1.77\phs\phm{0.00:}  &   17.20\phs\phm{0.00:}  &   51.05\phs\phm{0.00:}  & 2\\
NGC 5194                  & R\phm{ C} & 2.0 & S       & 13 27 46.9 &    47 27 16 &   640 &  7.7    &     4.0 & \phn 9.03 &   11.02\phs\phm{0.00:}  &   17.47\phs\phm{0.00:}  &  108.68\phs\phm{0.00:}  &  292.08\phs\phm{0.00:}  & 2\\
1335$+$39                 & \phm{R }C & 1.8 & S       & 13 35 28.3 &    39 24 33 &  6227 & \nodata & \nodata &     14.2  &    0.08$\pm$0.02\phm{:} &    0.15$\pm$0.02\phm{:} &    1.04$\pm$0.03\phm{:} &    2.40$\pm$0.09\phm{:} & 1\\
NGC 5252                  & \phm{R }C & 1.9 & A       & 13 35 44.0 &    04 47 43 &  7179 & \nodata &  $-$2.0 &     14.5  & $< 0.15$\phs\phm{0.00:} &    0.19$\pm$0.05\phm{:} &    0.43$\pm$0.05\phm{:} &    1.06$\pm$0.12\phm{:} & 1\\
NGC 5256                  & \phm{R }C & 2.0 & S       & 13 36 14.7 &    48 31 53 &  8521 & \nodata &    99.0 &     14.1  &    0.28$\pm$0.03\phm{:} &    1.13$\pm$0.04\phm{:} &    7.19$\pm$0.03\phm{:} &   10.35$\pm$0.16\phm{:} & 3\\
NGC 5283                  & \phm{R }C & 2.0 & A       & 13 39 40.7 &    67 55 33 &  2776 & 41.4    &  $-$2.0 &     14.3  & $< 0.07$\phs\phm{0.00:} & $< 0.06$\phs\phm{0.00:} &    0.19$\pm$0.03\phm{:} & $< 0.41$\phs\phm{0.00:} & 1\\
NGC 5273                  & R C       & 1.9 & A       & 13 39 55.2 &    35 54 18 &  1268 & 21.3    &  $-$2.0 &     12.71 &    0.12$\pm$0.03\phm{:} &    0.33$\pm$0.02\phm{:} &    1.08$\pm$0.04\phm{:} &    1.12$\pm$0.15\phm{:} & 1\\
Mrk 461                   & \phm{R }C & 2.0 & A       & 13 45 04.4 &    34 23 57 &  5071 & \nodata &     2.0 &     14.5  & $< 0.10$\phs\phm{0.00:} &    0.19$\pm$0.05\phm{:} &    0.34$\pm$0.04\phm{:} &    0.98$\pm$0.13:       & 1\\
NGC 5347                  & \phm{R }C & 2.0 & A       & 13 51 05.7 &    33 44 16 &  2595 & 36.7    &     2.0 &     13.18 &    0.38$\pm$0.02\phm{:} &    1.14$\pm$0.03\phm{:} &    1.40$\pm$0.03\phm{:} &    3.29$\pm$0.14\phm{:} & 1\\
Mrk 279\tablenotemark{a}  & \phm{R }C & 1.0 & S       & 13 51 51.9 &    69 33 13 &  9241 & \nodata &  $-$2.0 &     14.5  &    0.24$\pm$0.02\phm{:} &    0.50$\pm$0.02\phm{:} &    1.55$\pm$0.03\phm{:} &    2.38$\pm$0.06\phm{:} & 1\\
NGC 5395\tablenotemark{a} & R\phm{ C} & 2.0 & S       & 13 56 29.7 &    37 40 05 &  3682 & \nodata &     3.0 &     12.47 &    0.70$\pm$0.03\phm{:} &    1.17$\pm$0.02\phm{:} &    9.94$\pm$0.05\phm{:} &   19.28$\pm$0.07\phm{:} & 1\\
NGC 5548                  & R C       & 1.5 & A       & 14 15 44.0 &    25 22 01 &  5354 & \nodata &     0.0 &     13.54 &    0.42$\pm$0.03\phm{:} &    0.78$\pm$0.03\phm{:} &    1.08$\pm$0.05\phm{:} &    1.78$\pm$0.09\phm{:} & 1\\
Mrk 471                   & \phm{R }C & 1.8 & S       & 14 20 46.9 &    33 04 37 & 10450 & \nodata &     1.0 &     14.5  &    0.18$\pm$0.03\phm{:} &    0.11$\pm$0.03\phm{:} &    0.93$\pm$0.03\phm{:} &    2.33$\pm$0.09\phm{:} & 1\\
NGC 5631                  & R\phm{ C} & 2.0 & A       & 14 25 00.1 &    56 48 26 &  2082 & 32.7    &  $-$2.0 &     12.84 & $< 0.07$:\phs\phm{0.00} &    0.14$\pm$0.02\phm{:} &    0.42$\pm$0.04\phm{:} &    0.94$\pm$0.10\phm{:} & 1\\
NGC 5674                  & \phm{R }C & 1.9 & S       & 14 31 22.3 &    05 40 43 &  7701 & \nodata &     5.0 &     13.7  &    0.17$\pm$0.02\phm{:} &    0.36$\pm$0.04\phm{:} &    1.59$\pm$0.05\phm{:} &    3.70$\pm$0.16\phm{:} & 1\\
Mrk 817                   & \phm{R }C & 1.5 & S       & 14 34 58.0 &    59 00 40 &  9481 & \nodata & \nodata &     14.3  &    0.35$\pm$0.02\phm{:} &    1.40$\pm$0.02\phm{:} &    2.33$\pm$0.03\phm{:} &    2.29$\pm$0.06\phm{:} & 1\\
NGC 5695                  & \phm{R }C & 2.0 & A       & 14 35 20.7 &    36 47 13 &  4386 & \nodata &     1.0 &     13.9  & $< 0.09$\phs\phm{0.00:} &    0.16$\pm$0.02\phm{:} &    0.65$\pm$0.03\phm{:} &    1.74$\pm$0.09\phm{:} & 1\\             
Mrk 841                   & \phm{R }C & 1.5 & A       & 15 01 36.3 &    10 37 56 & 11105 & \nodata & \nodata &     14.0  &    0.19$\pm$0.03\phm{:} &    0.58$\pm$0.04\phm{:} &    0.51$\pm$0.02\phm{:} & $< 0.26$:\phs\phm{0.00} & 1\\
NGC 5929\tablenotemark{a} & \phm{R }C & 2.0 & S       & 15 24 18.3 &    41 50 43 &  2654 & 38.5    &     2.0 &     14.0  &    0.43$\pm$0.03\phm{:} &    1.62$\pm$0.03\phm{:} &    9.14$\pm$0.04\phm{:} &   13.69$\pm$0.12\phm{:} & 3\\
NGC 5940                  & \phm{R }C & 1.0 & S       & 15 28 51.1 &    07 37 38 & 10359 & \nodata &     2.0 &     14.3  &    0.15$\pm$0.03\phm{:} &    0.11$\pm$0.04\phm{:} &    0.79$\pm$0.03\phm{:} &    1.92$\pm$0.08\phm{:} & 1\\
NGC 6104                  & \phm{R }C & 1.5 & H       & 16 14 40.0 &    35 49 46 &  8437 & \nodata &     1.0 &     14.1  &    0.16$\pm$0.03\phm{:} &    0.19$\pm$0.02\phm{:} &    0.55$\pm$0.02\phm{:} &    1.76$\pm$0.07\phm{:} & 1\\
NGC 6951                  & R\phm{ C} & 2.0 & S       & 20 36 37.7 &    65 55 48 &  1280 & 24.1    &     4.0 &     12.41 &    1.36$\pm$0.01\phm{:} &    2.18$\pm$0.01\phm{:} &   17.23$\pm$0.03\phm{:} &   43.59$\pm$0.17\phm{:} & 4\\
2237$+$07                 & \phm{R }C & 1.8 & S       & 22 37 46.4 &    07 47 37 &  7132 & \nodata &     1.0 &     14.3  &    0.19$\pm$0.03\phm{:} &    0.53$\pm$0.02\phm{:} &    1.01$\pm$0.03\phm{:} &    0.97$\pm$0.20\phm{:} & 1\\
NGC 7469                  & \phm{R }C & 1.0 & S       & 23 00 44.4 &    08 36 19 &  4536 & \nodata &     1.0 &     13.0  &    1.60$\pm$0.05\phm{:} &    5.84$\pm$0.05\phm{:} &   27.68$\pm$0.04\phm{:} &   34.91$\pm$0.64\phm{:} & 3\\
NGC 7479                  & R\phm{ C} & 1.9 & S       & 23 02 26.8 &    12 03 06 &  2021 & 32.4    &     5.0 &     11.93 &    1.40$\pm$0.04\phm{:} &    3.92$\pm$0.07\phm{:} &   15.35$\pm$0.06\phm{:} &   24.60$\pm$0.31\phm{:} & 3\\
Mrk 530                   & \phm{R }C & 1.5 & S       & 23 16 22.7 & $-$00 01 48 &  8492 & \nodata &     3.0 &     14.4  &    0.32$\pm$0.04\phm{:} & $< 0.17$\phs\phm{0.00:} &    1.25$\pm$0.04\phm{:} &    2.03$\pm$0.09\phm{:} & 1\\
NGC 7674                  & \phm{R }C & 2.0 & S       & 23 25 24.4 &    08 30 06 &  8405 & \nodata &     4.0 &     13.6  &    0.68$\pm$0.04\phm{:} &    1.88$\pm$0.04\phm{:} &    5.28$\pm$0.05\phm{:} &    7.91$\pm$0.17\phm{:} & 3\\
NGC 7682                  & \phm{R }C & 2.0 & A       & 23 26 30.2 &    03 15 28 &  4762 & \nodata &     2.0 &     14.3  &    0.14$\pm$0.03\phm{:} &    0.29$\pm$0.04:       &    0.59$\pm$0.05\phm{:} &    1.07$\pm$0.22\phm{:} & 1\\
NGC 7743                  & R\phm{ C} & 2.0 & H       & 23 41 48.6 &    09 39 25 &  1307 & 24.4    &  $-$1.0 &     12.9  &    0.20$\pm$0.04\phm{:} &    0.20$\pm$0.03\phm{:} &    1.24$\pm$0.04\phm{:} &    4.93$\pm$0.19\phm{:} & 1\\
\enddata

\tablecomments{Col. (1): Object name given in Ho et al. (1997$a$) or Osterbrock \& Martel (1993). Col. (2): Sample (R = RSA; C = CfA). Col. (3): Classification based on the optical line emission. Col. (4): Classification based on the FIR continuum emission (A = AGN dominant; S = starburst dominant; H = host dominant). Cols. (5)--(6): Position in Equatorial coordinates (1950 equinox). Col. (7): Recession velocity in the cosmic microwave background frame. Col. (8): Distance taken from Tully (1988). Col. (9): Hubble type index. Col. (10): Magnitude on the Zwicky-$B(0)$ system. We have not studied the CfA Seyfert $1.0+1.5$ galaxies with $m_B > 14.0$. Cols. (11)--(14): {\it IRAS} flux densities at 12, 25, 60, and 100 \micron. The colon (:) indicates that the uncertainty might be greater than the face value. Col. (14): Reference for the $IRAS$ data (1 = this work; 2 = Rice et al. 1988; 3 = Soifer et al. 1989; 4 = Sanders et al. 1995).}  

\tablenotetext{a}{Member of an interacting system. The companion(s) might affect the {\it IRAS} data.}
\tablenotetext{b}{Not observed by the {\it IRAS}.}
\tablenotetext{c}{Non-interacting pair with NGC 4319, which contributes seriously to the {\it IRAS} data. Thus this objects has been excluded from our analyses.}

\end{deluxetable}


\begin{deluxetable}{lrrrrrlrrrrc}

\tabletypesize{\scriptsize}
\setlength{\tabcolsep}{0.025in}
\tablenum{4}
\tablecolumns{12}
\tablewidth{0pc}
\tablecaption{MRK STARBURST GALAXIES}

\tablehead{
\colhead{Name} &
\colhead{$\alpha$(1950)} &
\colhead{$\delta$(1950)} &
\colhead{$V$} &
\colhead{$D$} &
\colhead{$T$} &
\colhead{$m_B$} &
\colhead{$S_{12}$} &
\colhead{$S_{25}$} &
\colhead{$S_{60}$} &
\colhead{$S_{100}$} &
\colhead{Ref.} \\
\colhead{} &
\colhead{h m s} &
\colhead{d m s} &
\colhead{km/s} &
\colhead{Mpc} &
\colhead{} &
\colhead{mag} &
\colhead{Jy} &
\colhead{Jy} &
\colhead{Jy} &
\colhead{Jy} &
\colhead{} \\
\colhead{(1)} &
\colhead{(2)} &
\colhead{(3)} &
\colhead{(4)} &
\colhead{(5)} &
\colhead{(6)} &
\colhead{(7)} &
\colhead{(8)} &
\colhead{(9)} &
\colhead{(10)} &
\colhead{(11)} &
\colhead{(12)} }

\startdata
Mrk 13                     & 07 51 56.9 &    60 26 17 &  1506 & 24.9    &     3.0 & 14.5  &    0.13$\pm$0.02:       & $< 0.09$\phs\phm{0.00:} &    0.44$\pm$0.03\phm{:} &    1.42$\pm$0.22\phm{:} & 1 \\
Mrk 14                     & 08 05 21.7 &    72 56 33 &  3188 & \nodata & \nodata & 14.4  & $< 0.05$\phs\phm{0.00:} &    0.10$\pm$0.02\phm{:} &    0.93$\pm$0.03\phm{:} &    1.20$\pm$0.07\phm{:} & 1 \\
Mrk 25                     & 10 00 22.2 &    59 40 43 &  2776 & \nodata & \nodata & 14.2  &    0.14$\pm$0.03\phm{:} &    0.24$\pm$0.02\phm{:} &    1.36$\pm$0.03\phm{:} &    1.97$\pm$0.11\phm{:} & 1 \\
Mrk 52                     & 12 23 09.0 &    00 51 00 &  2488 & 33.5    &  $-$1.0 & 12.85 &    0.30$\pm$0.05\phm{:} &    1.25$\pm$0.07\phm{:} &    4.76$\pm$0.04\phm{:} &    6.09$\pm$0.13\phm{:} & 1 \\
Mrk 87                     & 08 15 55.2 &    74 08 53 &  2863 & 41.3    &     0.5 & 13.4  &    0.12$\pm$0.02\phm{:} &    0.39$\pm$0.02\phm{:} &    1.85$\pm$0.04\phm{:} &    3.75$\pm$0.07\phm{:} & 1 \\
Mrk 90                     & 08 26 15.7 &    52 51 53 &  4411 & \nodata & \nodata & 13.9  &    0.13$\pm$0.02\phm{:} &    0.15$\pm$0.03\phm{:} &    1.17$\pm$0.03\phm{:} &    2.12$\pm$0.08\phm{:} & 1 \\
Mrk 100                    & 08 54 29.8 &    66 39 47 &  3682 & \nodata &    99.0 & 14.2  & $< 0.08$\phs\phm{0.00:} &    0.06$\pm$0.01\phm{:} &    1.38$\pm$0.07\phm{:} &    2.96$\pm$0.46:       & 1 \\
Mrk 102                    & 09 08 18.1 &    46 50 33 &  4452 & \nodata &  $-$6.0 & 14.3  & $< 0.08$\phs\phm{0.00:} &    0.25$\pm$0.03\phm{:} &    0.72$\pm$0.04\phm{:} &    1.50$\pm$0.14:       & 1 \\
Mrk 133\tablenotemark{a}   & 09 57 52.0 &    72 21 53 &  2116 & 32.4    &     4.0 & 12.8  &    0.25$\pm$0.03\phm{:} &    0.67$\pm$0.02\phm{:} &    3.29$\pm$0.04\phm{:} &    5.58$\pm$0.18\phm{:} & 1 \\
Mrk 149                    & 10 34 38.9 &    64 31 32 &  1819 & 27.1    & \nodata & 14.4  & $< 0.11$\phs\phm{0.00:} &    0.20$\pm$0.02\phm{:} &    0.44$\pm$0.03\phm{:} &    0.52$\pm$0.06\phm{:} & 1 \\  
Mrk 158                    & 10 56 02.2 &    61 47 56 &  2265 & 33.0    &     1.0 & 13.0  &    0.35$\pm$0.02\phm{:} &    1.21$\pm$0.02\phm{:} &    8.55$\pm$0.03\phm{:} &   12.84$\pm$0.16\phm{:} & 2 \\
Mrk 161                    & 10 59 07.3 &    45 29 47 &  6212 & \nodata &    99.0 & 13.4  &    0.13$\pm$0.03\phm{:} &    0.38$\pm$0.02\phm{:} &    2.49$\pm$0.06\phm{:} &    4.10$\pm$0.11\phm{:} & 1 \\
Mrk 185                    & 11 38 36.0 &    47 58 13 &  3317 & \nodata &     6.0 & 13.0  &    0.21$\pm$0.02\phm{:} &    0.17$\pm$0.02\phm{:} &    2.49$\pm$0.03\phm{:} &    6.18$\pm$0.10\phm{:} & 1 \\
Mrk 190                    & 11 49 10.1 &    48 57 34 &  1186 & 17.0    &  $-$5.0 & 13.1  &    0.22$\pm$0.02\phm{:} &    0.40$\pm$0.03\phm{:} &    2.86$\pm$0.03\phm{:} &    5.03$\pm$0.10\phm{:} & 1 \\
Mrk 201                    & 12 11 41.3 &    54 48 19 &  2678 & 39.1    &    10.0 & 13.0  &    0.84$\pm$0.03\phm{:} &    4.57$\pm$0.03\phm{:} &   25.66$\pm$0.04\phm{:} &   26.21$\pm$0.13\phm{:} & 2 \\
Mrk 203\tablenotemark{b}   & 12 15 43.6 &    44 27 00 &  7623 & \nodata &     5.0 & 14.2  &    \nodata\phm{0...}    &    \nodata\phm{0...}    &    \nodata\phm{0...}    &    \nodata\phm{0...}    & \nodata \\
Mrk 213                    & 12 29 02.7 &    58 14 25 &  3264 & \nodata &     1.0 & 13.2  &    0.36$\pm$0.02\phm{:} &    0.59$\pm$0.03\phm{:} &    4.05$\pm$0.03\phm{:} &    6.27$\pm$0.12\phm{:} & 1 \\
Mrk 220\tablenotemark{a}   & 12 41 31.8 &    55 10 13 &  5037 & \nodata & \nodata & 14.5  &    0.09$\pm$0.03\phm{:} &    0.14$\pm$0.02\phm{:} &    1.75$\pm$0.05\phm{:} &    2.70$\pm$0.20\phm{:} & 1 \\
Mrk 286                    & 14 18 48.9 &    71 48 55 &  7592 & \nodata &    99.0 & 13.9  &    0.16$\pm$0.02\phm{:} &    0.76$\pm$0.03\phm{:} &    4.84$\pm$0.04\phm{:} &    6.92$\pm$0.09\phm{:} & 1 \\
Mrk 307                    & 22 33 31.4 &    20 03 53 &  5215 & \nodata &     5.0 & 13.7  &    0.37$\pm$0.02\phm{:} &    0.27$\pm$0.03:       &    1.83$\pm$0.03\phm{:} &    4.00$\pm$0.19\phm{:} & 1 \\
Mrk 319                    & 23 16 10.4 &    24 57 27 &  7757 & \nodata &     1.0 & 14.0  &    0.28$\pm$0.03\phm{:} &    0.41$\pm$0.03\phm{:} &    4.61$\pm$0.04\phm{:} &    7.49$\pm$0.33\phm{:} & 1 \\
Mrk 326                    & 23 25 36.0 &    23 15 17 &  3200 & \nodata &     3.5 & 13.9  &    0.20$\pm$0.04\phm{:} &    0.68$\pm$0.04\phm{:} &    3.77$\pm$0.05\phm{:} &    6.24$\pm$0.10\phm{:} & 1 \\
Mrk 353                    & 01 00 35.0 &    22 04 26 &  4358 & \nodata &     3.0 & 14.2  &    0.17$\pm$0.02\phm{:} &    0.47$\pm$0.04\phm{:} &    3.97$\pm$0.04\phm{:} &    5.37$\pm$0.14\phm{:} & 1 \\
Mrk 363                    & 01 48 12.0 &    21 45 00 &  2675 & \nodata &  $-$2.0 & 13.9  &    0.17$\pm$0.04\phm{:} &    0.23$\pm$0.05\phm{:} &    2.36$\pm$0.04\phm{:} &    3.40$\pm$0.39\phm{:} & 1 \\
Mrk 384                    & 08 00 08.4 &    23 32 00 &  4903 & \nodata &     3.0 & 14.2  &    0.27$\pm$0.03\phm{:} &    0.57$\pm$0.04\phm{:} &    4.12$\pm$0.04\phm{:} &    6.61$\pm$0.12\phm{:} & 1 \\
Mrk 391                    & 08 51 32.3 &    39 43 40 &  4180 & \nodata &     1.0 & 13.9  &    0.19$\pm$0.03\phm{:} &    0.32$\pm$0.04\phm{:} &    1.28$\pm$0.03\phm{:} &    3.23$\pm$0.10\phm{:} & 1 \\
Mrk 401                    & 09 27 19.6 &    29 45 33 &  1958 & 26.8    &     0.0 & 13.6  &    0.21$\pm$0.02\phm{:} &    0.61$\pm$0.04\phm{:} &    2.64$\pm$0.03\phm{:} &    3.62$\pm$0.08\phm{:} & 1 \\
Mrk 409                    & 09 46 44.7 &    32 27 26 &  1785 & 24.1    &  $-$2.0 & 14.2  & $< 0.08$\phs\phm{0.00:} & $< 0.10$\phs\phm{0.00:} &    0.32$\pm$0.03\phm{:} &    0.68$\pm$0.07\phm{:} & 1 \\
Mrk 430                    & 11 48 28.0 &    55 21 20 &  6026 & \nodata &     0.0 & 13.4  &    0.14$\pm$0.02\phm{:} &    0.30$\pm$0.02\phm{:} &    0.84$\pm$0.03\phm{:} &    1.61$\pm$0.10\phm{:} & 1 \\
Mrk 439                    & 12 22 07.8 &    39 39 33 &  1278 & 21.6    &     1.0 & 12.84 &    0.30$\pm$0.03\phm{:} &    0.73$\pm$0.05\phm{:} &    5.76$\pm$0.03\phm{:} &   11.30$\pm$0.15\phm{:} & 2 \\
Mrk 446                    & 12 47 44.0 &    33 25 47 &  7347 & \nodata &     3.0 & 14.2  &    0.27$\pm$0.04:       &    0.29$\pm$0.04\phm{:} &    1.48$\pm$0.04\phm{:} &    2.29$\pm$0.10\phm{:} & 1 \\
Mrk 449                    & 13 09 12.0 &    36 32 47 &  1358 & 22.8    &     1.0 & 13.5  &    0.11$\pm$0.03:       &    0.53$\pm$0.03\phm{:} &    2.42$\pm$0.03\phm{:} &    4.47$\pm$0.10\phm{:} & 1 \\
Mrk 480                    & 15 04 44.4 &    42 50 00 &  5516 & \nodata & \nodata & 14.2  &    0.08$\pm$0.02\phm{:} &    0.14$\pm$0.01\phm{:} &    1.83$\pm$0.03\phm{:} &    3.29$\pm$0.08\phm{:} & 1 \\
Mrk 489\tablenotemark{c}   & 15 42 36.2 &    41 14 33 &  9562 & \nodata & \nodata & 14.2  &    0.26$\pm$0.02\phm{:} &    0.56$\pm$0.02\phm{:} &    3.31$\pm$0.03\phm{:} &    7.05$\pm$0.07\phm{:} & 1 \\
Mrk 496                    & 16 10 24.0 &    52 35 00 &  8818 & \nodata & \nodata & 14.0  &    0.29$\pm$0.02\phm{:} &    1.22$\pm$0.03\phm{:} &    6.25$\pm$0.04\phm{:} &    9.34$\pm$0.10\phm{:} & 2 \\
Mrk 538                    & 23 33 41.2 &    01 52 42 &  2442 & 36.9    &     3.0 & 13.1  &    0.47$\pm$0.04\phm{:} &    2.82$\pm$0.09\phm{:} &   10.52$\pm$0.04\phm{:} &   11.66$\pm$0.12\phm{:} & 2 \\
Mrk 545                    & 00 07 18.6 &    25 38 42 &  4236 & \nodata &     1.0 & 12.5  &    0.59$\pm$0.04\phm{:} &    1.24$\pm$0.05\phm{:} &    8.77$\pm$0.05\phm{:} &   14.96$\pm$0.10\phm{:} & 2 \\
Mrk 575                    & 01 45 52.9 &    12 21 51 &  5202 & \nodata &     1.0 & 14.0  &    0.34$\pm$0.04\phm{:} &    1.08$\pm$0.06\phm{:} &    2.95$\pm$0.07\phm{:} &    4.91$\pm$0.71\phm{:} & 1 \\
Mrk 602                    & 02 57 14.1 &    02 34 24 &  2646 & 34.2    &     3.5 & 13.8  &    0.15$\pm$0.03\phm{:} &    0.74$\pm$0.03\phm{:} &    3.89$\pm$0.03\phm{:} &    5.94$\pm$0.09\phm{:} & 1 \\
Mrk 617                    & 04 31 35.5 & $-$08 40 42 &  4710 & \nodata &     5.0 & 14.0  &    1.44$\pm$0.03\phm{:} &    7.82$\pm$0.03\phm{:} &   33.12$\pm$0.09\phm{:} &   36.19$\pm$0.48\phm{:} & 2 \\
Mrk 691                    & 15 44 43.2 &    18 02 22 &  3423 & \nodata &     7.0 & 13.2  &    0.40$\pm$0.02\phm{:} &    0.51$\pm$0.02\phm{:} &    4.41$\pm$0.05\phm{:} &    8.06$\pm$0.09\phm{:} & 1 \\
Mrk 708                    & 09 39 34.5 &    04 54 07 &  2364 & \nodata & \nodata & 14.0  &    0.25$\pm$0.04\phm{:} &    0.84$\pm$0.04\phm{:} &    5.76$\pm$0.04\phm{:} &    8.69$\pm$0.27\phm{:} & 2 \\
Mrk 710                    & 09 52 10.2 &    09 30 32 &  1818 & \nodata &     2.0 & 13.5  &    0.17$\pm$0.03\phm{:} &    0.36$\pm$0.06\phm{:} &    2.95$\pm$0.03\phm{:} &    4.89$\pm$0.21\phm{:} & 1 \\
Mrk 743                    & 11 35 37.8 &    12 23 20 &  1328 & 17.0    &  $-$2.0 & 13.34 & $< 0.09$\phs\phm{0.00:} &    0.20$\pm$0.04\phm{:} &    1.42$\pm$0.04\phm{:} &    2.85$\pm$0.08:       & 1 \\
Mrk 752\tablenotemark{a}   & 11 50 10.0 &    02 01 03 &  6479 & \nodata &     4.0 & 14.4  & $< 0.08$\phs\phm{0.00:} &    0.30$\pm$0.05:       &    0.81$\pm$0.04\phm{:} &    1.21$\pm$0.12\phm{:} & 1 \\
Mrk 759                    & 12 08 04.6 &    16 18 42 &  2488 & 34.5    &     5.0 & 13.25 &    0.31$\pm$0.03\phm{:} &    0.43$\pm$0.05\phm{:} &    4.39$\pm$0.02\phm{:} &    9.26$\pm$0.10\phm{:} & 1 \\
Mrk 769                    & 12 22 53.9 &    16 44 49 &  2027 & 16.8    &     1.0 & 12.90 &    0.32$\pm$0.05\phm{:} &    1.06$\pm$0.06\phm{:} &    8.40$\pm$0.04\phm{:} &   12.69$\pm$0.14\phm{:} & 2 \\
Mrk 781                    & 12 51 19.6 &    09 58 49 &  3145 & \nodata &     4.0 & 13.5  &    0.33$\pm$0.04\phm{:} &    0.46$\pm$0.07\phm{:} &    2.11$\pm$0.04\phm{:} &    4.27$\pm$0.10\phm{:} & 1 \\
Mrk 799                    & 13 59 08.5 &    59 34 10 &  3134 & \nodata &     3.0 & 13.08 &    0.65$\pm$0.03\phm{:} &    1.92$\pm$0.03\phm{:} &   10.82$\pm$0.04\phm{:} &   21.51$\pm$0.07\phm{:} & 2 \\
Mrk 809                    & 14 20 10.2 &    13 56 40 &  7901 & \nodata &     8.0 & 14.5  &    0.22$\pm$0.04\phm{:} &    0.18$\pm$0.04\phm{:} &    2.07$\pm$0.04\phm{:} &    3.39$\pm$0.09\phm{:} & 1 \\
Mrk 1002                   & 01 34 41.1 &    05 37 23 &  2872 & \nodata &  $-$2.0 & 13.5  &    0.45$\pm$0.04\phm{:} &    0.69$\pm$0.06\phm{:} &    5.07$\pm$0.06\phm{:} &    6.47$\pm$0.16\phm{:} & 1 \\
Mrk 1194                   & 05 09 06.6 &    05 08 26 &  4459 & \nodata &  $-$2.0 & 13.7  &    0.38$\pm$0.02\phm{:} &    0.84$\pm$0.03\phm{:} &    6.93$\pm$0.06\phm{:} &   11.56$\pm$0.17\phm{:} & 1 \\ 
Mrk 1301                   & 11 33 10.8 &    35 36 44 &  1874 & \nodata &     0.0 & 14.5  & $< 0.14$\phs\phm{0.00:} &    0.77$\pm$0.07\phm{:} &    1.72$\pm$0.06\phm{:} &    2.45$\pm$0.23\phm{:} & 1 \\
Mrk 1344                   & 13 06 41.8 & $-$05 00 24 &  3498 & \nodata &  $-$2.0 & 14.5  &    0.17$\pm$0.04\phm{:} &    0.68$\pm$0.08\phm{:} &    3.15$\pm$0.07\phm{:} &    2.49$\pm$0.15\phm{:} & 1 \\
NGC 2782                   & 09 10 54.1 &    40 19 18 &  2770 & 37.3    &     1.0 & 12.66 &    0.71$\pm$0.04\phm{:} &    1.58$\pm$0.05\phm{:} &    9.60$\pm$0.05\phm{:} &   14.65$\pm$0.18\phm{:} & 2 \\
NGC 3395\tablenotemark{a}  & 10 47 02.3 &    33 14 45 &  1899 & 27.4    &     6.0 & 12.1  &    0.43$\pm$0.03\phm{:} &    1.42$\pm$0.05\phm{:} &   10.77$\pm$0.03\phm{:} &   17.46$\pm$0.11\phm{:} & 2 \\
NGC 3504                   & 11 00 28.1 &    28 14 35 &  1837 & 26.5    &     2.0 & 11.80 &    1.13$\pm$0.03\phm{:} &    4.21$\pm$0.06\phm{:} &   22.70$\pm$0.06\phm{:} &   35.70$\pm$0.13\phm{:} & 2 \\
\enddata

\clearpage

\tablecomments{Col. (1): Object name given in Balzano (1983). Cols. (2)--(3): Position in Equatorial coordinates (1950 equinox). Col. (4): Recession velocity in the cosmic microwave background frame. Col. (5): Distance taken from Tully (1988). Col. (6): Hubble type index. Col. (7): Magnitude on the Zwicky-$B(0)$ system. Cols. (8)--(11): {\it IRAS} flux densities at 12, 25, 60, and 100 \micron. The colon (:) indicates that the uncertainty might be greater than the face value. Col. (12): Reference for the $IRAS$ data (1 = this work; 2 = Soifer et al. 1989).}  

\tablenotetext{a}{Member of an interacting system. The companion(s) might affect the {\it IRAS} data.}
\tablenotetext{b}{Not observed by the {\it IRAS}.}
\tablenotetext{c}{Non-interacting pair with a galaxy that might affect the {\it IRAS} data.}

\end{deluxetable}


\begin{deluxetable}{lcccccc}

\tabletypesize{\scriptsize}
\setlength{\tabcolsep}{0.04in}
\tablenum{5}
\tablecolumns{7}
\tablewidth{0pc}
\tablecaption{MEDIAN VALUES}

\tablehead{
\colhead{Sample} &
\colhead{$\log(F_{\rm FIR}/F_B)$} &
\colhead{$\log(S_{100}/S_{60})$} &
\colhead{$\log(S_{60}/S_{25})$} &
\colhead{$\log(S_{25}/S_{12})$} &
\colhead{$\log(\nu_{12}S_{12}/F_{\rm FIR})$} &
\colhead{$C_{60}$} \\ }

\startdata
RSA Seyfert   &    $-0.240^{+0.317}_{-0.543}$ (50) & $0.425^{+0.117}_{-0.131}$ (47) & $0.773^{+0.129}_{-0.285}$ (46) & $0.180^{+0.158}_{-0.122}$ (45) & $-0.416^{+0.200}_{-0.094}$ (45) & $0.151^{+0.479}_{-0.481}$ (46) \\
CfA Seyfert   & \phs$0.014^{+0.342}_{-0.178}$ (45) & $0.242^{+0.131}_{-0.141}$ (45) & $0.502^{+0.241}_{-0.213}$ (45) & $0.351^{+0.136}_{-0.113}$ (44) & $-0.196^{+0.213}_{-0.304}$ (44) & $0.412^{+0.646}_{-0.766}$ (45) \\
              &    $-0.026^{+0.337}_{-0.167}$ (37) & $0.251^{+0.121}_{-0.150}$ (37) & $0.458^{+0.232}_{-0.176}$ (37) & $0.371^{+0.103}_{-0.098}$ (36) & $-0.243^{+0.267}_{-0.285}$ (36) & $0.343^{+0.515}_{-0.697}$ (37) \\
Mrk starburst & \phs$0.320^{+0.189}_{-0.179}$ (56) & $0.211^{+0.072}_{-0.050}$ (56) & $0.807^{+0.091}_{-0.142}$ (56) & $0.386^{+0.156}_{-0.183}$ (55) & $-0.553^{+0.139}_{-0.097}$ (56) & $1.189^{+0.237}_{-0.466}$ (56) \\
\enddata

\tablecomments{Differences to the 25$^{th}$ and 75$^{th}$ percentiles are also given. The object numbers are given in parentheses. The spectral curvature at 60 \micron\ is defined as $C_{60} = \log(S_{100}/S_{60})/\log(\nu_{100}/\nu_{60}) - \log(S_{60}/S_{25})/\log(\nu_{60}/\nu_{25})$. For CfA Seyfert galaxies, the upper-row values have been obtained from the entire sample. The lower-row values have been obtained by excluding type $1.0+1.5$ objects with $m_B > 14.0$. For Mrk starburst galaxies, the values have been obtained from objects with $m_B \le 14.5$.}
\end{deluxetable}

\clearpage
\section*{FIGURE CAPTIONS}

\figcaption[Fig1]{$F_{\rm FIR}/F_{B}$ ($a$) and $S_{100}/S_{60}$ ($b$) for Mrk starburst galaxies. The abscissa is the Hubble type index $T$. The symbol ``U'' indicates that the $T$ value is unavailable. The FIR emission of an object above the dotted line is dominated by the starburst ($a$) or host galaxy ($b$). We plot all the galaxies where the {\it IRAS} data are available. \label{Fig1}}

\figcaption[Fig2]{Number distribution of the morphological type for RSA Seyfert galaxies. ($a$) AGN-dominant objects. ($b$) starburst-dominant objects. ($c$) host-dominant objects. {\it Open areas}: Seyfert $1.0 + 1.5$. {\it Hatched areas}: Seyfert $1.8 + 1.9$. {\it Filled areas}: Seyfert 2.0. The bins along the abscissa have the following means: ``E'' = E, ``S0'' = S0, ``Sa'' = S0/a--Sab, ``Sb'' = Sb--Sbc, ``Sc'' = Sc--Scd, ``Sd'' = Sd--Sdm, ``Sm'' = Sm--Im, and ``Pec'' = Pec. We plot all the galaxies. The distributions of AGN- and starburst-dominant objects are different with the confidence level 99.8\% \label{Fig2}}

\figcaption[Fig3]{Number distribution of the morphological type for CfA Seyfert galaxies. ($a$) AGN-dominant objects. ($b$) starburst-dominant objects. ($c$) host-dominant objects. {\it Open areas}: Seyfert $1.0 + 1.5$. {\it Hatched areas}: Seyfert $1.8 + 1.9$. {\it Filled areas}: Seyfert 2.0. The abscissa is the same as in Fig. 2. We plot all the galaxies, except for $1335+39$ and Mrk 841 where the morphological type is unavailable as well as Seyfert $1.0+1.5$ galaxies with $m_B > 14.0$. The distributions of AGN- and starburst-dominant objects are different with the confidence level 98.2\%. \label{Fig3}}

\figcaption[Fig4]{Number distribution of the narrow-line Balmer decrement $F_{{\rm H}\alpha}/F_{{\rm H}\beta}$ for RSA Seyfert galaxies. ($a$) AGN-dominant objects. ($b$) starburst-dominant objects. ($c$) host-dominant objects. {\it Open areas}: Seyfert $1.0 + 1.5$. {\it Hatched areas}: Seyfert $1.8 + 1.9$. {\it Filled areas}: Seyfert 2.0. We plot all the galaxies, except for NGC 4579 where the $F_{{\rm H}\alpha}/F_{{\rm H}\beta}$ value is unavailable. The distributions of AGN- and starburst-dominant objects are different with the confidence level 99.8\%. \label{Fig4}}

\figcaption[Fig5]{Number distribution of the flux density ratio $S_{100}/S_{60}$ for galaxies with $F_{\rm FIR}/F_B > 1$. ($a$) RSA Seyfert galaxies. ($b$) CfA Seyfert galaxies. ($c$) Mrk starburst galaxies. {\it Open areas}: Seyfert $1.0 + 1.5$. {\it Hatched areas}: Seyfert $1.8 + 1.9$. {\it Filled areas}: Seyfert 2.0 or starburst. We plot all the galaxies, except for CfA Seyfert $1.0+1.5$ galaxies with $m_B > 14.0$. The distributions of RSA and Mrk galaxies are different with the confidence level 98.8\%. Those of CfA and Mrk galaxies are different with the confidence level 95.7\%. \label{Fig5}}

\figcaption[Fig6]{{\it Upper panel}: Number distribution of the interaction class {\it IAC} for CfA Seyfert galaxies. ($a$) AGN-dominant objects. ($b$) starburst-dominant objects. {\it Lower panel}: Number distribution of the correlation function amplitude $\xi_0$ in units of Mpc$^{1.77}$ for CfA Seyfert galaxies. ($c$) AGN-dominant objects. ($d$) starburst-dominant objects. {\it Open areas}: Seyfert $1.0 + 1.5$. {\it Hatched areas}: Seyfert $1.8 + 1.9$. {\it Filled areas}: Seyfert 2.0. In ($a$), we plot Mrk 335, Mrk 573, $0152+06$ (UM 146), NGC 863, NGC 3516, NGC 4051, NGC 4151, NGC 4235, NGC 5283, NGC 5273, NGC 5548, NGC 5695, and Mrk 841. In ($b$), we plot Mrk 334, NGC 1068, NGC 1144, NGC 3227, NGC 4253, NGC 5033, NGC 5256, Mrk 471, NGC 5929, NGC 7469, and NGC 7674. In ($c$), we plot A$1058+45$ (UGC 6100), NGC 3516, NGC 4235, NGC 5252, NGC 5283, Mrk 461, NGC 5347, NGC 5548, NGC 5695, and Mrk 841. In ($d$), we plot NGC 4253, NGC 4388, $1335+39$ (UGC 8621), NGC 5256, Mrk 471, NGC 5674, and NGC 5929. The {\it IAC} distributions between AGN- and starburst-dominant objects are different with the confidence level 97.2\%. The $\xi_0$ distributions are different with the confidence level 84.7\%. \label{Fig6}}

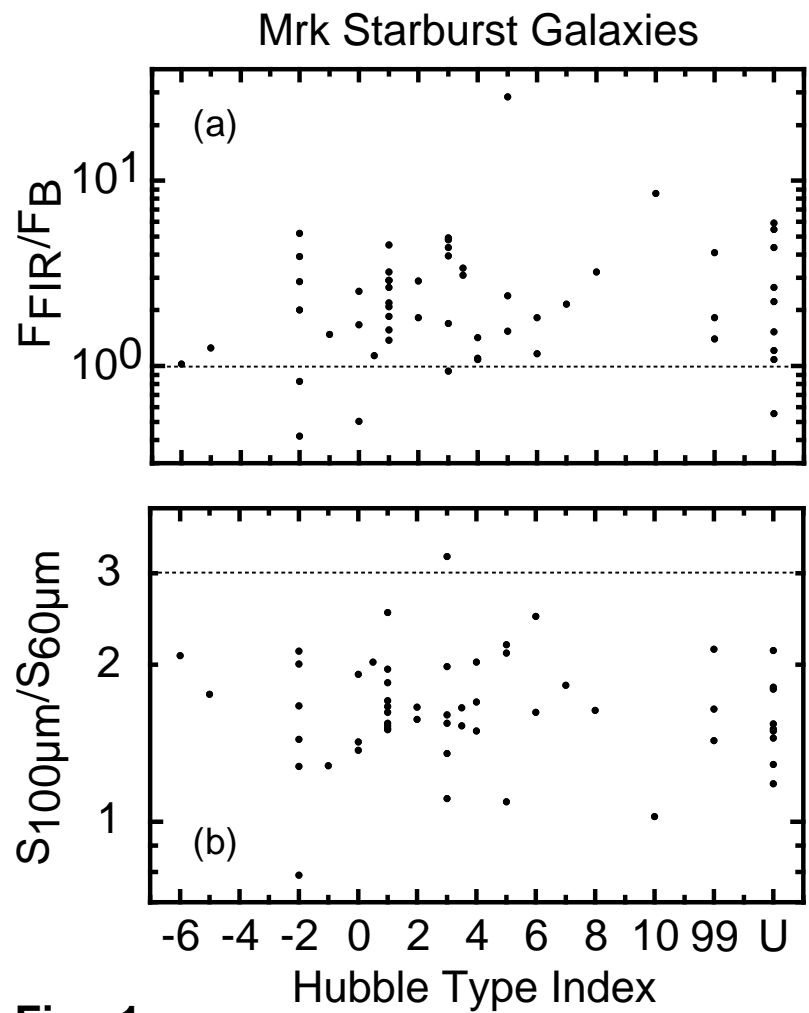
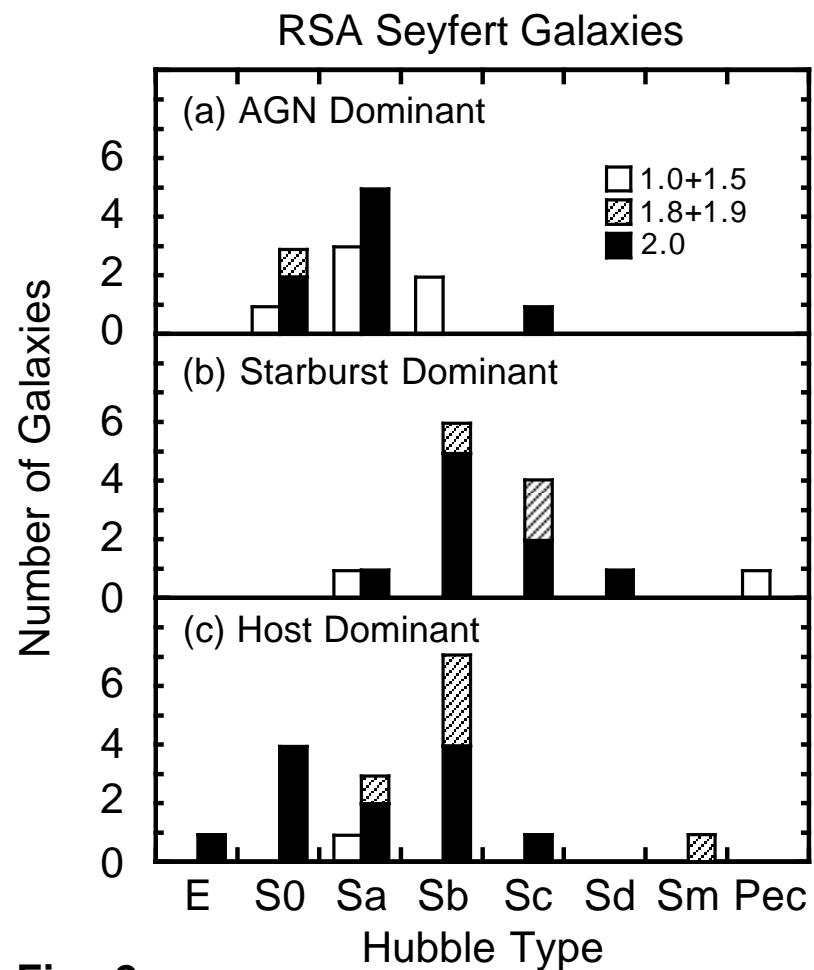
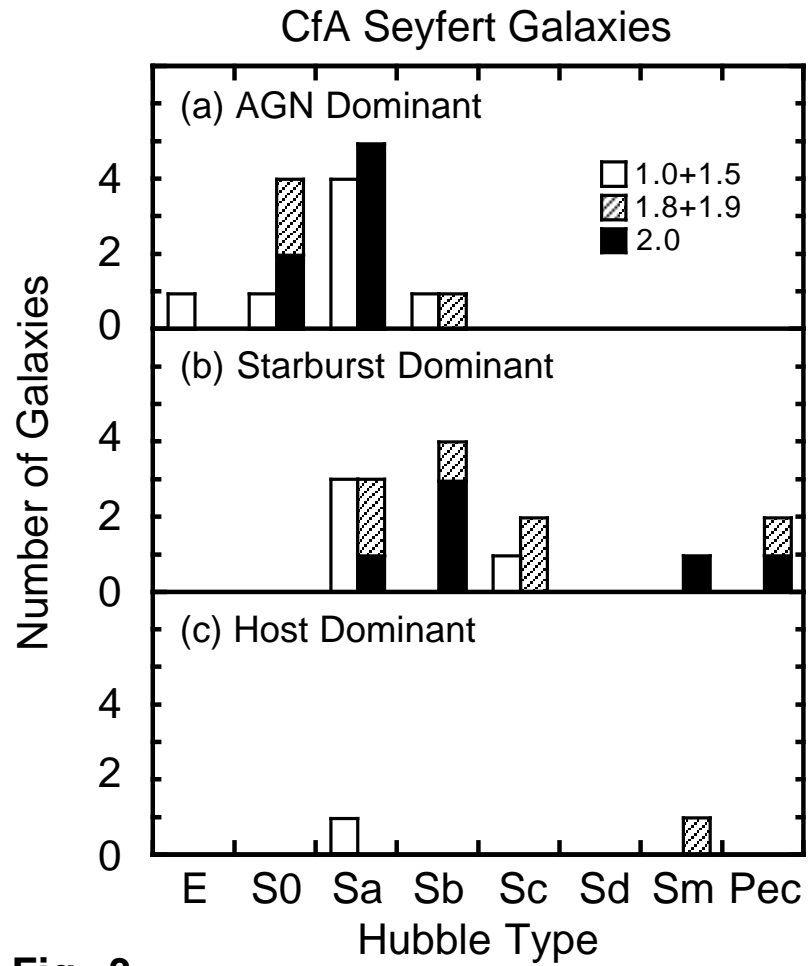
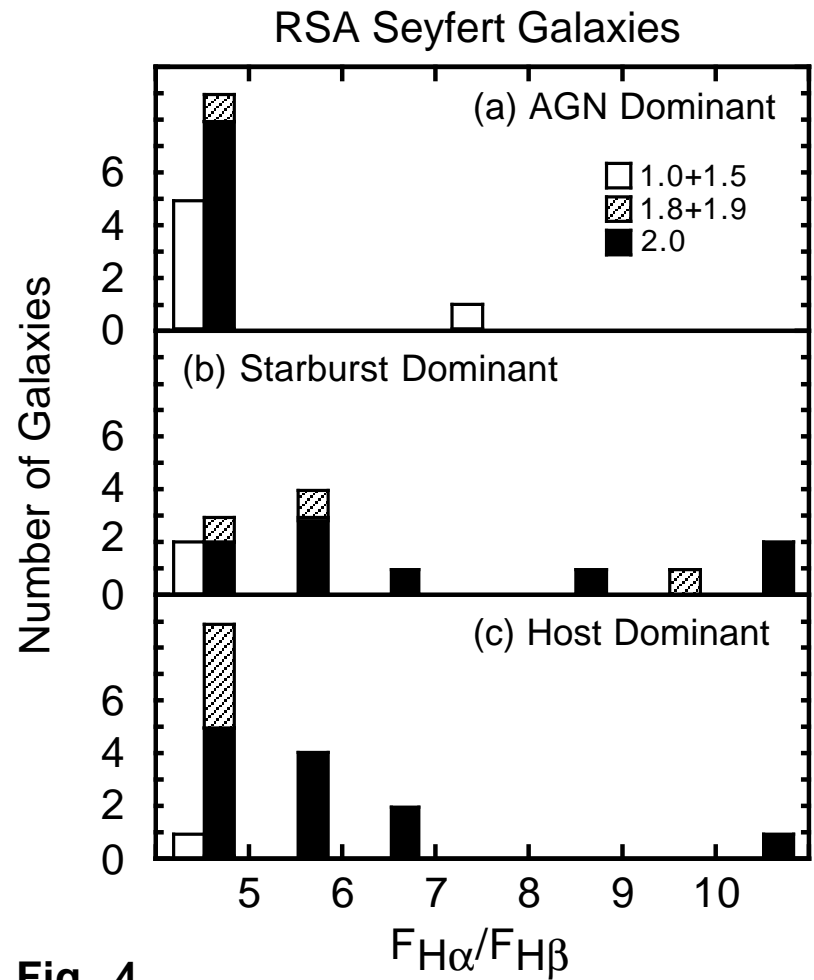
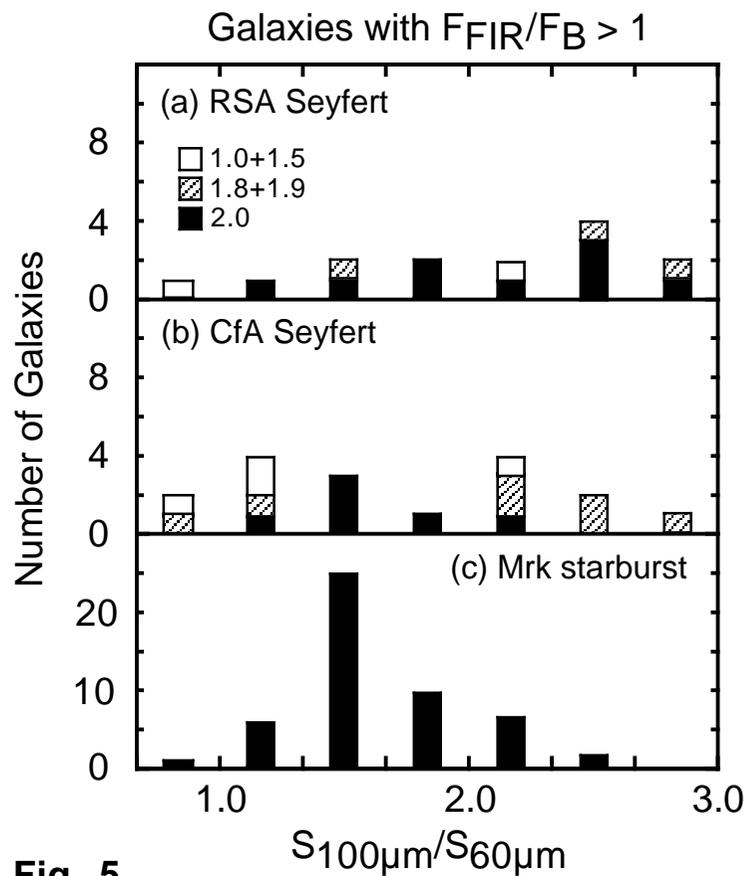
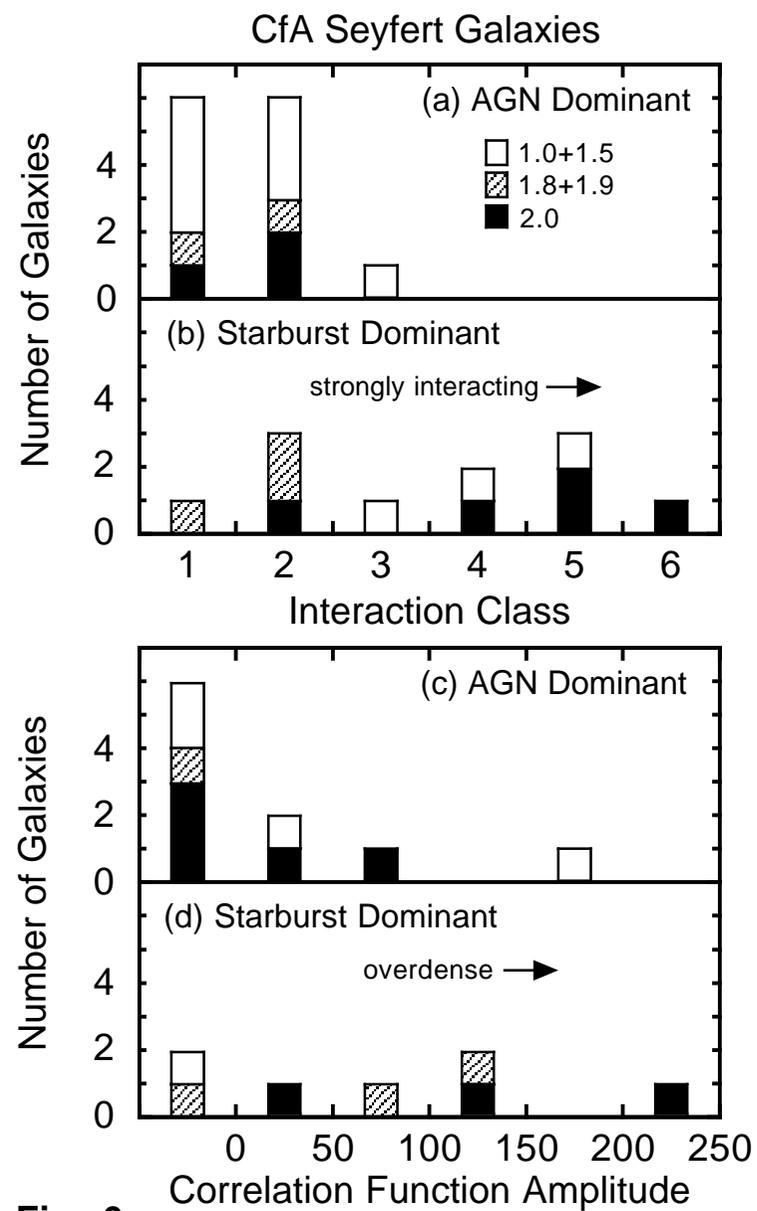
\end{document}